\documentclass[twocolumn]{aastex63}%manuscript

\newcommand\kepler{\emph{Kepler}}
\newcommand\kt{\emph{K2}}
\newcommand\gaia{\emph{Gaia}}

\usepackage{hyperref}
\usepackage{xcolor}
\usepackage{amssymb}
\usepackage{amsmath}

\accepted{\today}
\shorttitle{Scaling \kt. I. Stars}
\shortauthors{Hardegree-Ullman et al.}

\begin{document}

\title{Scaling \kt. I. Revised Parameters for 222,088 \kt\ Stars and a \kt\ Planet Radius Valley at 1.9 $R_{\oplus}$}

\correspondingauthor{Kevin K.\ Hardegree-Ullman}
\email{kevinkhu@caltech.edu}

\author[0000-0003-3702-0382]{Kevin K.\ Hardegree-Ullman}
\affiliation{Caltech/IPAC-NASA Exoplanet Science Institute, Pasadena, CA 91125}

\author[0000-0003-1848-2063]{Jon K.\ Zink}
\affiliation{Department of Physics and Astronomy, University of California, Los Angeles, CA 90095}
\affiliation{Caltech/IPAC-NASA Exoplanet Science Institute, Pasadena, CA 91125}

\author[0000-0002-8035-4778]{Jessie L.\ Christiansen}
\affiliation{Caltech/IPAC-NASA Exoplanet Science Institute, Pasadena, CA 91125}

\author[0000-0001-8189-0233]{Courtney D.\ Dressing}
\affiliation{Astronomy Department, University of California, Berkeley, CA 94720}

\author[0000-0002-5741-3047]{David R.\ Ciardi}
\affiliation{Caltech/IPAC-NASA Exoplanet Science Institute, Pasadena, CA 91125}

\author[0000-0001-5347-7062]{Joshua E.\ Schlieder}
\affiliation{Exoplanets and Stellar Astrophysics Laboratory, Code 667, NASA Goddard Space Flight Center, Greenbelt, MD 20771}

\begin{abstract}

Previous measurements of stellar properties for \kt\ stars in the Ecliptic Plane Input Catalog \citep[EPIC;][]{Huber2016} largely relied on photometry and proper motion measurements, with some added information from available spectra and parallaxes. Combining \gaia\ DR2 distances with spectroscopic measurements of effective temperatures, surface gravities, and metallicities from the Large Sky Area Multi-Object Fibre Spectroscopic Telescope (LAMOST) DR5, we computed updated stellar radii and masses for 26,838 \kt\ stars. For 195,250 targets without a LAMOST spectrum, we derived stellar parameters using random forest regression on photometric colors trained on the LAMOST sample. In total, we measured spectral types, effective temperatures, surface gravities, metallicities, radii, and masses for 222,088 A, F, G, K, and M-type \kt\ stars. With these new stellar radii, we performed a simple reanalysis of 299 confirmed and 517 candidate \kt\ planet radii from Campaigns 1--13, elucidating a distinct planet radius valley around $1.9\,R_{\oplus}$, a feature thus far only conclusively identified with \kepler\ planets, and tentatively identified with \kt\ planets. These updated stellar parameters are a crucial step in the process toward computing \kt\ planet occurrence rates.

\end{abstract}


\section{Introduction} \label{sec:intro}
The ubiquity of exoplanets in the Galaxy has been established by NASA's \kepler\ Telescope \citep{Borucki2010}, with the discovery of thousands of confirmed and candidate planets\footnote{\href{https://exoplanetarchive.ipac.caltech.edu/docs/counts\_detail.html}{https://exoplanetarchive.ipac.caltech.edu/docs/counts\_detail.html}} in both the \kepler\ prime and subsequent \kt\ missions. After the failure of two reaction wheels on \kepler, the \kt\ mission was commissioned, which allowed the \kepler\ spacecraft to stare at different fields along the ecliptic plane for approximately 80 days at a time, using radiation pressure from the Sun to act as a third stabilization axis \citep{Howell2014}. 

Our knowledge of the hundreds of confirmed and candidate planets discovered in the \kt\ data relies on accurate and precise stellar radius measurements for their host stars. In large surveys of hundreds of thousands of stars, like \kt, it is practical to rely on stellar properties derived from readily available data. The values for \kt\ targets in the Ecliptic Planet Input Catalog (EPIC) come from \citet{Huber2016}, which were measured with \textsf{galclassify}\footnote{\href{https://github.com/danxhuber/galclassify}{https://github.com/danxhuber/galclassify}}, which uses the \emph{Galaxia} synthetic Milky Way model \citep{Sharma2011} and the Padova isochrones \citep{Girardi2000,Marigo2007,Marigo2008}. The input sources to \textsf{galclassify} were reduced proper motions, spectra from the Large Sky Area Multi-Object Fiber Spectroscopic Telescope DR1 \citep[LAMOST;][]{Luo2015}, the Radial Velocity Experiment DR4 \citep[RAVE;][]{Kordopatis2013}, and Apache Point Observatory Galactic Evolution Experiment DR12 \citep[APOGEE;][]{Alam2015}, parallax measurements from \emph{Hipparcos} \citep{Vanleeuwen2007}, and photometric measurements from the US Naval Observatory CCD Astrograph Catalog \citep[UCAC4;][]{Zacharias2013}, the Sloan Digital Sky Survey \citep[SDSS;][]{Skrutskie2006}, and the Two Micron All Sky Survey \citep[2MASS;][]{Skrutskie2006}. For \kt\ Campaigns 1--8, 81\% of the stars were characterized using colors and reduced proper motions, 11\% from colors only, 7\% from spectroscopy, and 1\% from parallaxes and colors \citep{Huber2016}.

Since the EPIC was released, the European Space Agency's \gaia\ mission \citep{Gaia2016} has now measured parallaxes for over 1.3 billion sources in DR2 \citep{Gaia2018}. Subsequently, \citet{Berger2018} revised the radii of \kepler\ stars and planets, reducing typical uncertainties on those measurements by a factor of 4--5 in most cases. Measurements of stellar parameters in the EPIC were largely based on photometry and proper motions, which can introduce biases in derived properties like temperature and surface gravity. \citet{Huber2016} noted specifically for subgiants that 55\%--70\% were misclassified as dwarf stars. Consequently, stellar properties for these stars had large uncertainties. Since the different \kt\ fields span a wide range of galactic latitudes, these biases are potentially caused by poor measurements of interstellar extinction. Additionally, the Padova isochrones are known to underestimate the radii of cool stars \citep{Boyajian2012}, and \citet{Huber2016} caution that EPIC M dwarf radii can be underestimated by up to 20\%. The exquisite precision of the \gaia\ measurements, improved interstellar extinction maps such as those from \citet{Green2018}, and recent empirical calibrations for cool stars \citep{Mann2015,Mann2019}, allow us to better constrain absolute magnitudes and refine stellar parameters based on photometry.

A moderate resolution stellar spectrum can be used to constrain basic stellar parameters more precisely than photometry alone, such as spectral type, effective temperature ($T_{\mathrm{eff}})$, surface gravity ($\log\,g$), and metallicity, which is commonly measured as iron abundance [Fe/H]. For transiting exoplanet studies, planet radius measurements are limited by the precision to which we know the radius of their host star. With bolometric luminosities and effective temperatures we can measure stellar radii ($R_{\star}$) from the Stefan-Boltzmann law. If surface gravity is also constrained, then a stellar mass ($M_{\star}$) can also be measured, which is necessary for constraining planet masses from radial velocities.

Several catalogs of \kt\ planets have gathered spectra of planet candidate host stars \citep[e.g.,][]{Crossfield2016,Dressing2017,Martinez2017,Dressing2017b,Petigura2018,Mayo2018,Dressing2019}. Different instruments and analysis techniques, however, produce different results, necessitating cross calibration between catalogs if conclusions are to be drawn about planet populations across the \kt\ campaigns. Stars without known or candidate planets are often overlooked for spectroscopic stellar characterization. This information is needed for accurate studies of planet occurrence rate calculations by spectral type, and drawing conclusions about planet host and non-host populations. Of course, photometry is much more readily available than spectroscopy for most stars, but large spectroscopic surveys such as LAMOST, RAVE, and APOGEE provide a wealth of information for millions of stars which are unbiased toward planet hosts.

Precise stellar radii for planet hosts can also reveal information about underlying planet populations. Indeed, one of the key results from the \kepler\ mission was the discovery of a planet radius valley between $\sim$1.5 and 2.0 Earth radii ($R_{\oplus}$) by \citet{Fulton2017}, which was enabled by improved precision in stellar radius measurements from California-\kepler\ Survey spectra. This planet radius gap was independently observed using a smaller set of \kepler\ targets with stellar properties measured from asteroseismology \citep{Vaneylen2018}. The astrophysical origin of this effect has been explored by \citet{Owen2013}, \citet{Lee2014}, \citet{Lee2016}, \citet{Owen2017}, and \citet{Lopez2018}. Using \kt\ data, \citet{Mayo2018} and \citet{Kruse2019} both identified a `tentative' planet radius gap with their catalogs of 275 planet candidates from Campaigns 0--10 and 818 planet candidates from Campaigns 0--8, respectively. \citet{Mayo2018} computed stellar radii using \textsf{isochrones} \citep{Morton2015}, with inputs of effective temperature, surface gravity, and metallicity derived from high resolution ($R\approx44,000$) Tillinghast Reflector Echelle Spectrograph (TRES) optical spectra (5059--5317\,\AA). They compared their planet radius distribution to the \citet{Fulton2017} distribution, but found that a log-uniform distribution fit their data equally well, which they attribute to their relatively small planet sample. \citet{Kruse2019} used stellar radii from \gaia\ for 648 of their targets and from the EPIC for most of the remaining stars without a \gaia\ measurement. They also conservatively call their planet radius gap tentative due to planet radius uncertainties and a limited sample.

\begin{figure*}[ht]
    \gridline{\fig{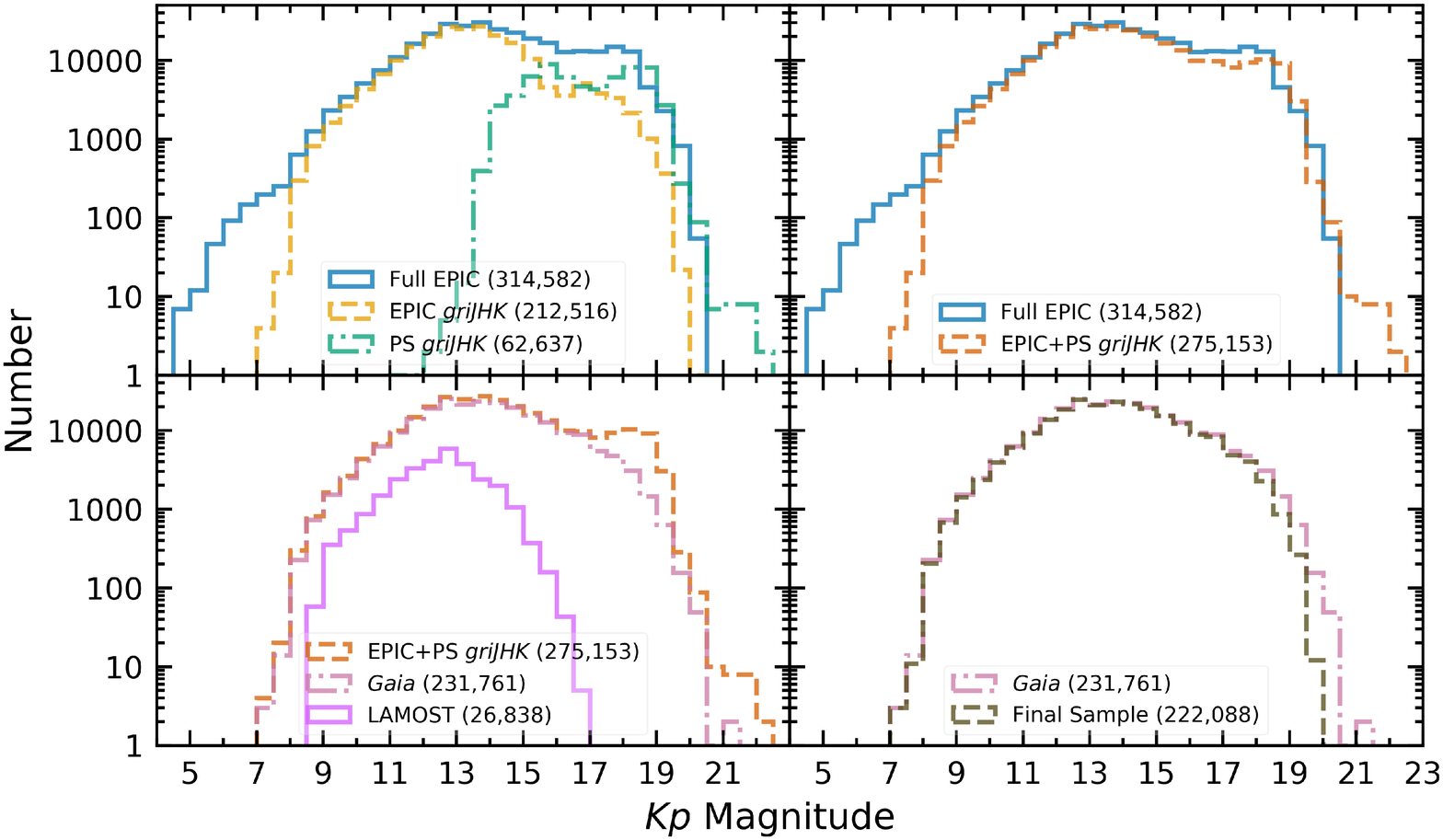}{1.0\textwidth}{(a)}}\vspace{-0.5em}
    \caption{Magnitude distributions highlight each of our sample cuts. (Upper left) The full EPIC catalog (solid), EPIC targets with full optical and 2MASS infrared photometry (dashed), and additional \kt\ targets with Pan-STARRS and 2MASS photometry (dashed dotted). $Kp$-band magnitudes were recomputed for \kt\ targets with Pan-STARRS photometry, which is why there are a few targets fainter than the original EPIC catalog. (Upper right) The full EPIC catalog (solid) and the combined EPIC and Pan-STARRS targets (dashed). (Lower left) The combined EPIC and Pan-STARRS targets (dashed), \kt\ targets with a \gaia\ parallax (dashed dotted), and targets with a LAMOST spectrum (solid). \gaia\ is nearly complete between $G=12$ and $G=17$ \citep{Gaia2018}, which explains why \gaia\ targets diminish beyond $Kp\approx17$. Understandably, targets with a LAMOST spectrum are relatively bright due to our S/N cuts. (Lower right) The \gaia\ (dashed dotted) and final target samples (dashed). Our color cuts mostly removed the faintest targets from our final sample.}\label{fig:1}
\end{figure*}

In this paper we leverage parallaxes from \gaia, stellar properties from LAMOST spectra, and photometry from the EPIC to calculate revised stellar properties (spectral type, distance, $T_{\mathrm{eff}}$, $\log\,g$, [Fe/H], $R_{\star}$, and $M_{\star}$) for 222,088 \kt\ stars. In Section~\ref{sec:catalog} we update target photometry and describe our target selection criteria from the EPIC, \gaia, and LAMOST. For stars with both a \gaia\ parallax and a LAMOST spectrum, we describe our spectroscopic stellar classification for A, F, G, and K (AFGK) type stars in Section~\ref{sec:afgk} and M dwarfs in Section~\ref{sec:mdwarfs}. We compute stellar properties for the remaining stars with only \gaia\ parallaxes and photometry in Section~\ref{sec:photclass}, In Section~\ref{sec:discussion}, we compare our revised stellar parameters to the EPIC, and remeasure \kt\ planet radii which we use to identify a clear \kt\ planet radius valley at $1.9\,R_{\oplus}$.

\section{Catalog} \label{sec:catalog}

We started with the \kt\ observed target catalog\footnote{\href{https://exoplanetarchive.ipac.caltech.edu/cgi-bin/TblView/nph-tblView?app=ExoTbls\&config=k2targets}{https://exoplanetarchive.ipac.caltech.edu/cgi-bin/TblView/nph-tblView?app=ExoTbls\&config=k2targets}}, which contains 342,964 targets with an object type of `star'. Several targets were observed in multiple campaigns, in which case we remove duplicate EPIC IDs, leaving us with 314,582 unique targets. Of these unique targets, there are 212,516 with UCAC4 or SDSS $g$, $r$, $i$, and 2MASS $J$, $H$, and $K_s$-band photometry, which we use later for target selection and stellar classification. Figure~\ref{fig:1} shows \kepler\ $Kp$-band magnitude distributions from the full EPIC catalog along with distributions from each of our target sample cuts, which we discuss in the following sections.

\subsection{Pan-STARRS Photometry} \label{sec:panstarrs}

There are 87,828 targets with complete $J$, $H$, and $K_s$-band photometry but incomplete or missing $g$, $r$, and $i$-band photometry. Using the EPIC IDs for these targets, we queried the Panoramic Survey Telescope and Rapid Response System \citep[Pan-STARRS;][]{Chambers2016} DR2 database \citep{Flewelling2016}. This resulted in $g$, $r$, and $i$-band photometry (mean PSF magnitudes) for 62,637 targets. These targets are on average between 2 and 2.5 magnitudes fainter than the EPIC targets with previous $g$, $r$, and $i$-band photometry (Figure~\ref{fig:1}), which is likely why they did not have previous optical measurements.

The average Pan-STARRS photometric uncertainties are about 10 times smaller than the average EPIC photometric uncertainties in the $g$ and $r$-bands, and comparable in $i$-band. Thus, we queried the Pan-STARRS database for all EPIC targets with previous optical measurements, resulting in 84,176 additional Pan-STARRS measurements. We use Pan-STARRS photometry for any of our targets fainter than the saturation limit ($g\lesssim14.5$\footnote{\href{https://outerspace.stsci.edu/display/PANSTARRS/PS1+FAQ+-+Frequently+asked+questions}{https://outerspace.stsci.edu/display/PANSTARRS/PS1+FAQ+-+Frequently+asked+questions}}; 123,819 targets), and the EPIC values otherwise. In total, we have 275,153 unique targets with complete $g$, $r$, $i$, $J$, $H$, and $K_s$-band photometry (Figure~\ref{fig:1}).

We recomputed the \kepler\ $Kp$ magnitude for all targets using our updated $g$, $r$, and $i$-band photometry and the following equations from \citet{Brown2011}:
\begin{subequations}
\begin{eqnarray}
    Kp = 0.25g + 0.75r, (g - r) \leq 0.3 \\
    Kp = 0.3g + 0.7i, (g - r) > 0.3
\end{eqnarray}
\end{subequations}

\begin{figure}[ht]
    \centering
    \includegraphics[width=0.46\textwidth]{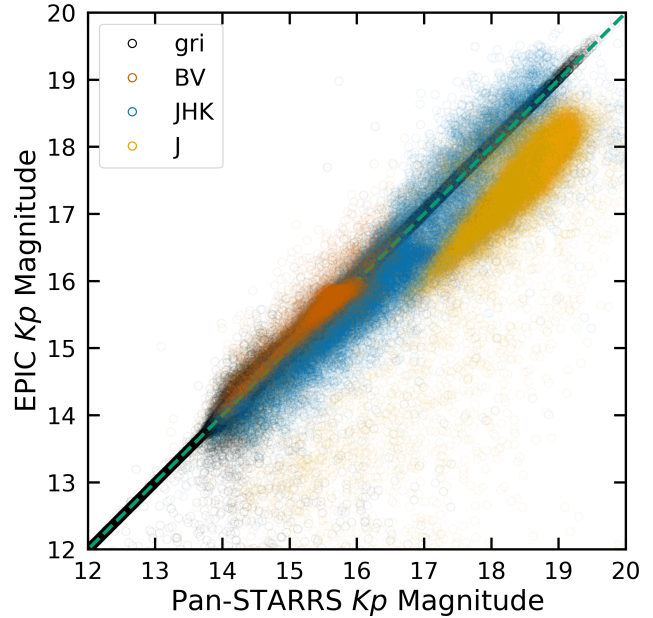}
    \caption{Comparison of $Kp$ magnitudes computed using Pan-STARRS $g$, $r$, and $i$-band photometry to previous measurements with optical (gri, black; BV, red) and infrared (JHK, blue; J, orange) photometry. Optical magnitude estimates are very similar. Previous measurements of $Kp$ from JHK photometry are skewed toward overestimating brightness. Previous $J$-band $Kp$ magnitude estimates are on average one magnitude brighter than our new optical measurements. The apparent truncation of Pan-STARRS $Kp$ measurements brighter than $\sim$14th magnitude for targets with previous estimates from BV, JHK, and $J$-band photometry is due to the Pan-STARRS saturation limit of $g\lesssim14.5$. For targets with Pan-STARRS measurements brighter than the saturation limit, we instead used EPIC gri photometry when available. We truncate the plot at 12th magnitude to highlight computed $Kp$ differences, and because brighter targets follow the one-to-one line (gray dashed).} \label{fig:2}  
		\vspace{-0.5em}
\end{figure}

Previous measurements of $Kp$ magnitudes were computed with less precise relationships from \citet{Brown2011} and \citet{Howell2012} using $B$, $V$, $J$, $H$, and $K_s$ photometry if $g$, $r$, and $i$-band photometry was unavailable \citep{Huber2016}. The \kepler\ $Kp$ filter response function ($\gtrsim 20\%$ transmission 4300--8900\,\AA\footnote{\url{https://keplergo.arc.nasa.gov/CalibrationResponse.shtml}}) overlaps with the $g$, $r$, and $i$-bands, so estimated magnitudes from these bands takes priority. We compared the newly computed $Kp$ magnitudes to previous estimates in Figure~\ref{fig:2}. Estimates from $J$-band photometry alone tend to yield $Kp$ measurements one magnitude brighter than from optical photometry.

\subsection{Gaia}
\label{sec:gaia}

\begin{figure*}[ht]
    \gridline{\fig{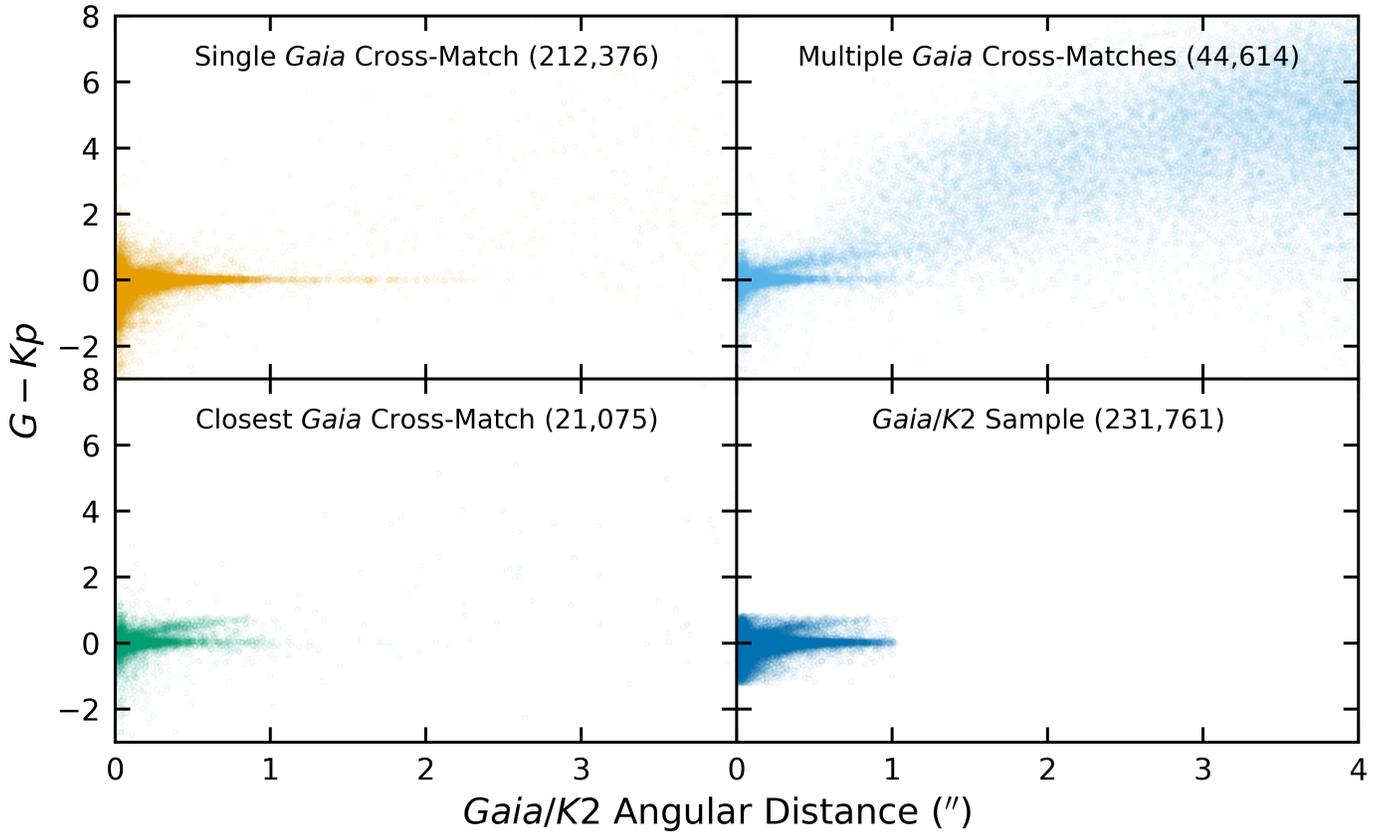}{1.0\textwidth}{(a)}}\vspace{-0.5em}
    \caption{$G-Kp$ versus \kt/\gaia\ angular distance for single \gaia\ cross-match sources (upper left) and multiple \gaia\ cross-matches (upper right). For the sources with multiple cross-matches, we selected the target with the closest distance to the origin, which effectively removed the multiple cross-match cloud (lower left). For the final sample, we selected targets within $3\sigma$ of the average angular distance and $G-Kp$ (lower right).}\label{fig:3}
\end{figure*}

We used the \gaia/\kt\ cross-match database\footnote{\href{http://gaia-kepler.fun/}{http://gaia-kepler.fun/}} to obtain distances to our \kt\ stars from \citet{Bailer-Jones2018}. The 4\arcsec\ radius cross-match between the aforementioned \kt\ observed star catalog and the \gaia\ DR2 catalog yields 361,488 \gaia/\kt\ entries and 294,114 unique EPIC IDs. We combined this cross-match table with our photometry table, reducing the \gaia/\kt\ cross-match sample to 256,990 \gaia\ sources within 4\arcsec\ of our 275,153 \kt\ targets. The \kt\ targets without a \gaia\ cross-match are on average $\sim$2.5 magnitudes fainter ($Kp=16.38$) than those with a cross-match ($Kp=13.85$), and about 60\% of these targets are likely giant stars \citep[based on $J-K$ versus $r-J$ colors;][]{Muirhead2015}.

The similarities between the \gaia\ $G$-band \citep[$\gtrsim 20\%$ transmission 4000--9000\,\AA;][]{Evans2018} and \kepler\ $Kp$-band helped us to identify our \kt\ target in the \gaia\ data in the case of multiple cross-matches, which could be a binary companion or background source. There are 212,376 \kt\ targets with a single \gaia\ cross-match within 4\arcsec, 21,075 \kt\ targets with more than one cross-match, and 41,702 without any \gaia\ matches. There are a total of 44,614 different \gaia\ IDs for the 21,075 \kt\ targets with more than one cross-match.

We plot $G-Kp$ versus \kt/\gaia\ angular distance in Figure~\ref{fig:3} for both single and multiple cross-matches. If there were multiple cross-matches, we selected the target closest to the origin in $G-Kp$ and angular distance space (21,075 targets). For the multiple cross-matches, the distribution roughly follows that of single cross-match targets, but with a distinct branch extending into a cloud of sources with $G-Kp \gtrsim 0$ and angular distance $\gtrsim 0\farcs75$. A simple investigation of targets along the extra branch in the closest \gaia\ cross-match plot does not indicate that these stars are distinct from the other closest match stars (e.g. common proper motion binary versus background star). Further analysis of this feature is encouraged but is beyond the scope of this work. For quality control, we selected targets from the single and closest \gaia\ cross-match lists within $3\sigma$ of the average angular distance ($\sim$1\arcsec) and $|G-Kp|$ ($\sim$1), leaving 231,761 unique targets (Figure~\ref{fig:1}).

\subsection{LAMOST Spectra} \label{sec:lamost}

LAMOST has a 4,000 fiber multi-object spectrograph (3690--9100\,\AA, $R\approx1,800$) to survey stars and galaxies in the northern hemisphere \citep{Cui2012}. LAMOST DR5 v3 contains over nine million\footnote{\href{http://dr5.lamost.org/}{http://dr5.lamost.org/}} spectra. The LAMOST DR5 AFGK type star catalog\footnote{\href{http://dr5.lamost.org/catalogue}{http://dr5.lamost.org/catalogue}} is comprised of 5,348,712 spectra across all evolutionary stages, and the M dwarf catalog contains 534,393 spectra. We chose to use only LAMOST spectra because it contains more spectra than either APOGEE or RAVE. This also mitigated any effects from cross-calibrating spectroscopic parameters from other surveys.

We selected AFGK spectra with signal-to-noise (S/N) $> 50$ in $g$ and $r$ bands, and M spectra with S/N $> 50$ in $r$ and $i$ bands. Additionally, for comparison to our \kt\ catalog, we required that the LAMOST targets also have associated $g$, $r$, and $i$ band photometry. Thousands of targets, as identified by their 2MASS designation, were observed more than once, in which case we kept the target with the highest S/N in the $r$ band. This left us with 1,440,423 AFGK and 50,158 M star spectra with a unique 2MASS designation.

We used the Centre de Données astronomiques de Strasbourg (CDS) cross-match service\footnote{\href{http://cdsxmatch.u-strasbg.fr/xmatch}{http://cdsxmatch.u-strasbg.fr/xmatch}} to cross-match our \gaia/\kt\ and LAMOST catalogs using a 4\arcsec\ search radius, yielding 29,134 AFGK and 1,737 M star matches. To ensure we matched the correct target, we checked that the absolute difference between $g$, $r$, and $i$ magnitudes in the LAMOST and EPIC catalogs were less than 0.15, a conservative $2\sigma$ from the median difference in each band. This left us with 25,450 AFGK and 1,388 M stars that are \kt\ targets with a LAMOST spectrum and \gaia\ parallax (Figure~\ref{fig:1}). For these targets, we computed absolute magnitudes for the $g, r, i, J, H,$ and $K_s$ band photometry from the EPIC catalog (Table~\ref{tab:1}), accounting for interstellar extinction using \textsf{dustmaps} \citep{Green2018}.

The LAMOST pipeline \citep{Luo2012,Luo2015} assigns a Morgan-Keenan spectral type to each spectrum. For the AFGK catalog, $T_{\mathrm{eff}}$, $\log g$, and [Fe/H] were determined from the LAMOST stellar parameters pipeline \citep{Wu2011}, which uses the University of Lyon Spectroscopic analysis Software (\textsf{ULySS}) spectrum fitting package \citep{Koleva2009}. For M dwarfs, spectral type and atomic and molecular line indices were determined using \textsf{The Hammer} \citep{Covey2007}, but other stellar parameters were not derived \citep{Yi2014}. We discuss derivation of stellar radii and masses for AFGK stars in Section~\ref{sec:afgk}. In Section~\ref{sec:mdwarfs} we compute $T_{\mathrm{eff}}$, $\log g$, [Fe/H], $R_{\star}$, and $M_{\star}$ for M dwarfs.

\section{AFGK Stellar Parameters} \label{sec:afgk}

Since the LAMOST pipeline provides $T_{\mathrm{eff}}$, $\log g$, and [Fe/H] for AFGK stars, we can readily compute stellar radii in a similar fashion to \citet{Fulton2018}. We first computed bolometric magnitudes ($M_{\mathrm{bol}}$) from $K_s$ band measurements, since $K_s$ is less affected by interstellar extinction than the other optical and near-infrared photometric bands:
\begin{equation}
    M_{\mathrm{bol}} = m_{K_s} - 5[\log_{10}(d)-1] - A_{K_s} - BC,
\end{equation}
where $d$ is the distance computed from \gaia\ parallax measurements \citep{Bailer-Jones2018}, $A_{K_s}$ is the $K_s$ band interstellar extinction computed using \textsf{dustmaps} \citep{Green2018}, and BC is the bolometric correction. Bolometric corrections were computed using \textsf{isoclassify}, which interpolates the Modules for Experiments in Stellar Astrophysics (MESA) Isochrones and Stellar Tracks (MIST) grid \citep{Dotter2016} over $T_{\mathrm{eff}}$, $\log g$, [Fe/H], and $A_{K_s}$. Bolometric luminosity ($L_{\mathrm{bol}}$) was calculated from bolometric magnitudes using:
\begin{equation}
    L_{\mathrm{bol}} = L_0 10^{-0.4M_{\mathrm{bol}}},
\end{equation}
where $L_0\equiv3.0128\times10^{28}\,\mathrm{W}$ \citep{Mamajek2015}. Finally, we computed $R_{\star}$ from the Stefan-Boltzmann law:
\begin{equation}
    R_{\star} = \left( \frac{L_{\mathrm{bol}}}{4\pi\sigma_{\mathrm{SB}}T_{\mathrm{eff}}^4} \right)^{1/2},
\end{equation}
where $\sigma_{\mathrm{SB}}$ is the Stefan-Boltzmann constant. Since we have both $R_{\star}$ and $\log g$ measurements, $M_{\star}$ was computed using $M_{\star} = 10^{\log g} \times R_{\star}^2/G$, where $G$ is the gravitational constant.

Uncertainties for parameters in this paper were computed using a Monte Carlo approach. For targets with symmetric uncertainties, we drew $10^4$ samples from a Gaussian distribution for each measured value and associated uncertainty. For targets with asymmetric uncertainties we drew $10^4$ samples from a split normal distribution, combining the left and right sides of two Gaussian distributions centered on the measured value and the negative and positive uncertainties. We propagated these distributions through each equation and took the median of the resultant distribution as the measured value and the 15.87 and 84.13 percentiles as the uncertainties. The average uncertainties on $R_{\star}$ and $M_{\star}$ for AFGK stars with LAMOST spectra are 4.4\% and 14.9\%, respectively. The very low uncertainties on these measurements are due to the $\sim$1\% uncertainties on $T_\mathrm{eff}$ and $\log\,g$ provided by the LAMOST pipeline for high S/N targets.

\section{M Dwarf Parameters} \label{sec:mdwarfs}

\subsection{Spectral Type}\label{sec:spt}

\begin{figure}[ht]
    \centering
    \includegraphics[width=0.46\textwidth]{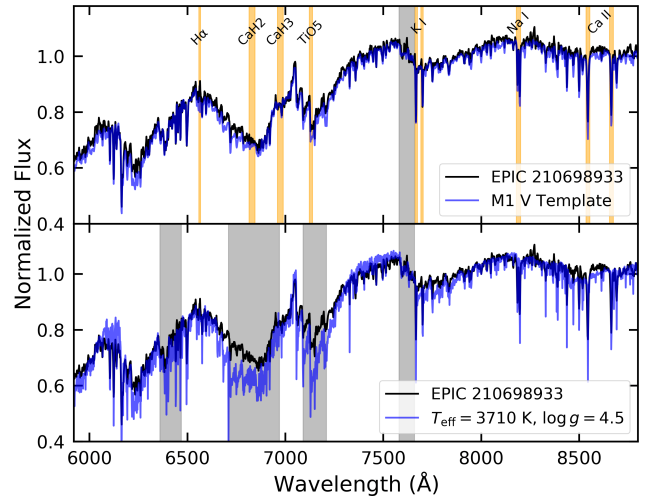}
    \caption{(Top) LAMOST spectrum of EPIC 210698933 (black) compared to an M1 V spectral template from \citet{Kesseli2017} (blue). A few prominent M dwarf atomic lines and molecular indices are indicated by the orange regions. (Bottom) The same spectrum compared to the closest matching PHOENIX-ACES model. Regions that the templates or models poorly matched the LAMOST spectra were masked out in the fitting process, which we show here in gray.} \label{fig:4}  
		\vspace{-0.5em}
\end{figure}

\begin{figure}[ht]
    \centering
    \includegraphics[width=0.46\textwidth]{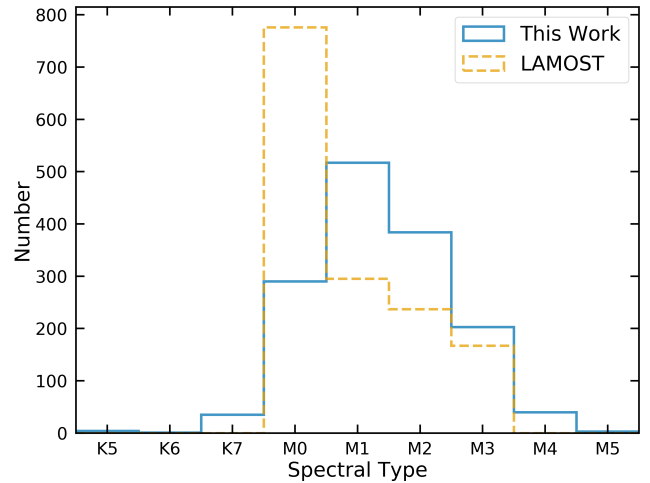}
    \caption{Comparison of spectral type classifications between LAMOST (dashed) and our pipeline (solid).} \label{fig:5}  
		\vspace{-0.5em}
\end{figure}

The LAMOST data do not include $T_{\mathrm{eff}}$, $\log g$, and [Fe/H] for M dwarfs, so we derived our own parameters for these stars. M dwarfs in the LAMOST catalog were initially classified using a modified version of \textsf{The Hammer} \citep{Covey2007}, then they were visually inspected, which changed the classification of nearly 1/5 of the stars \citep{Yi2014}. Since visual inspection can introduce bias, we re-spectral typed our LAMOST M dwarfs in a uniform automated process using the spectral templates of \citet{Kesseli2017}. These templates were derived from thousands of SDSS Baryon Oscillation Spectroscopic Survey (BOSS) spectra, covering 3600--10400\,\AA\ at a resolution of $R\approx2,000$ \citep{Dawson2013}. We used the K5 to M7 dwarf templates, which are separated into 0.5 dex metallicity bins. We resampled the template spectra to match the resolution of the LAMOST spectra using \textsf{SpectRes}\footnote{\href{https://github.com/ACCarnall/spectres}{https://github.com/ACCarnall/spectres}} \citep{Carnall2017}. To identify the closest matching spectral template, we minimize the goodness-of-fit statistic $G_K$ \citep[Equation 1 of][]{Cushing2008}, which is similar to $\chi^2$ minimization. In order to identify regions where the templates poorly fit our spectra, we ran the spectrum matching twice. First, using the same methods described in Section 5.1 of \citet{Mann2013b}, we computed the residuals from the best-fit spectral template for each of the LAMOST spectra, then computed the median fractional deviation between the data and the templates at each wavelength. Regions with a median deviation greater than 10\% were given a weight of 0 in $G_K$ for the second round of spectrum matching. This applied to $\lambda < 5910\,\textrm{\AA}$, $7580\,\textrm{\AA} < \lambda < 7660\,\textrm{\AA}$, and $\lambda > 8800\,\textrm{\AA}$. The poor fit at blue wavelengths might be due to the nature of the LAMOST spectra, which are taken in two different channels \citep[3700--5900\,\AA\ and 5700--9000\,\AA;][]{Cui2012} and combined during processing. Spectral typing of M dwarfs has historically been done at red wavelengths \citep[e.g.,][]{Kirkpatrick1991}, so we made no additional attempt to fit the red and blue regions separately. In the top panel of Figure~\ref{fig:4} we show an example M dwarf spectrum compared to its closest matching spectral template, with prominent atomic lines (H$\mathrm{\alpha}$, K I, Na I, Ca II) and molecular indices (CaH2, CaH3, TiO5) identified. In Figure~\ref{fig:5}, we compare our spectral types to those from the LAMOST pipeline for the same targets. Our classifications are more evenly distributed among the early M types with a peak near M1, whereas the LAMOST spectral types are significantly skewed toward M0. About 97\% of our targets are within one spectral type of the LAMOST classification. We also identified a few late K dwarf interlopers that were assigned an M spectral type by LAMOST. We derived parameters for these K dwarfs in the same manner as our spectroscopic M dwarfs described below.

\subsection{Effective Temperatures}\label{sec:teff}
We compared the LAMOST spectra to the PHOENIX-ACES model grid from \citet{Husser2013}, which were sampled in increments of $T_{\mathrm{eff}}=100\,\mathrm{K}$, $\log g = 0.5$, and $\mathrm{[Fe/H]} = 0.5$. From these model spectra, we interpolated a finer model grid to $T_{\mathrm{eff}}=10\,\mathrm{K}$ and $\log g = 0.1$, using $\mathrm{[Fe/H]} = 0$ models. To identify the closest matching model spectrum, we used the same procedure outlined in Section~\ref{sec:spt}, this time masking out the following regions: $\lambda<5920\,\textrm{\AA}$, $6360\,\textrm{\AA}<\lambda<6470\,\textrm{\AA}$, $6710\,\textrm{\AA}<\lambda<6970\,\textrm{\AA}$, $7090\,\textrm{\AA}<\lambda<7210\,\textrm{\AA}$, $7580\,\textrm{\AA}<\lambda<7660\,\textrm{\AA}$, and $\lambda>8800\,\textrm{\AA}$. In the bottom panel of Figure~\ref{fig:4}, we show the example spectrum compared to its closest matching spectral model, indicating regions that were masked out due to poor model fits. We adopt the temperatures from the closest matching spectral model but we refine our surface gravity measurements in Section~\ref{sec:radmassgrav}.

\begin{figure}[ht!]
\gridline{\fig{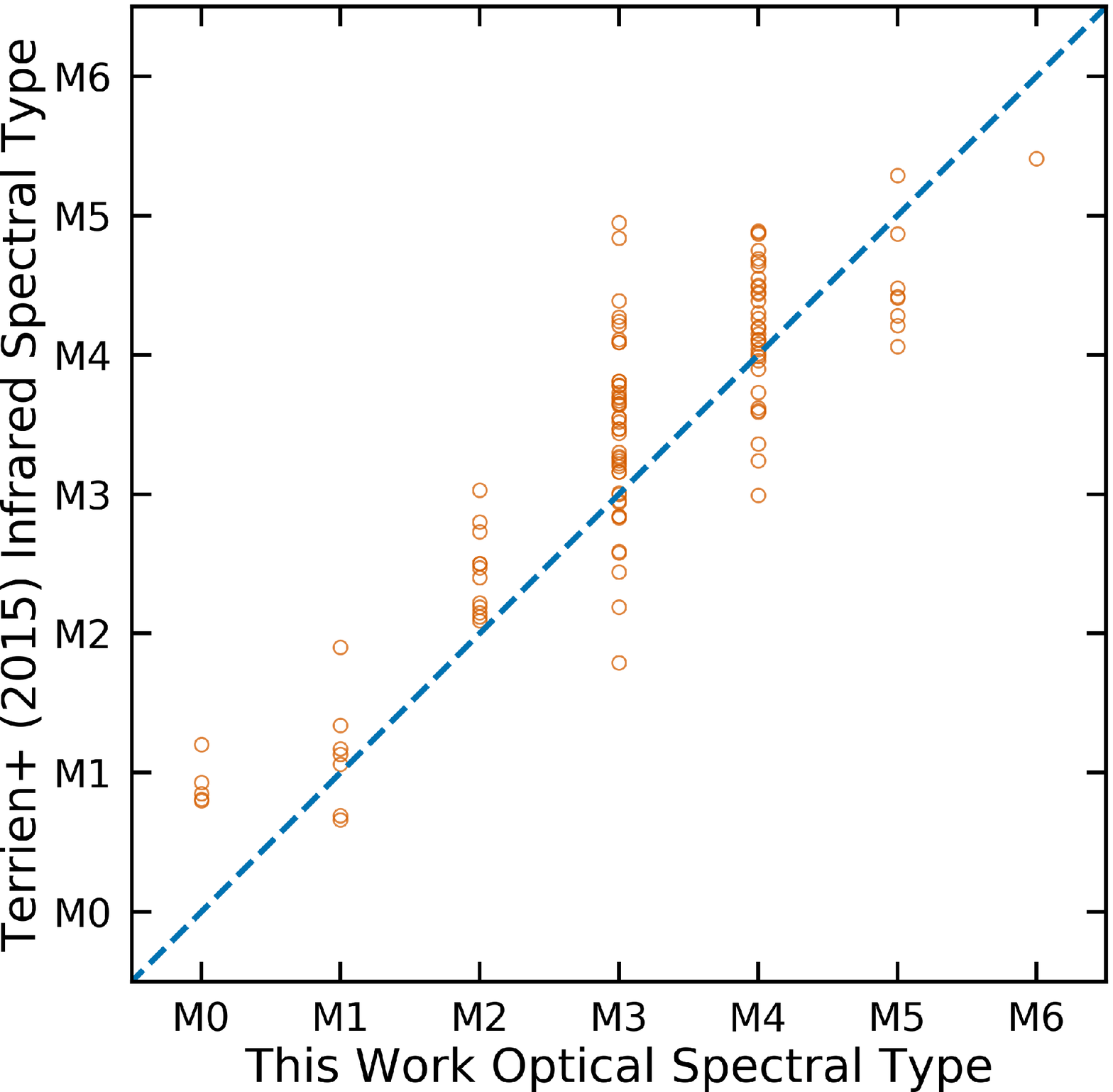}{0.445\textwidth}{\ \ \ \ \ \ \ \ \ \ (a)}}
\gridline{\fig{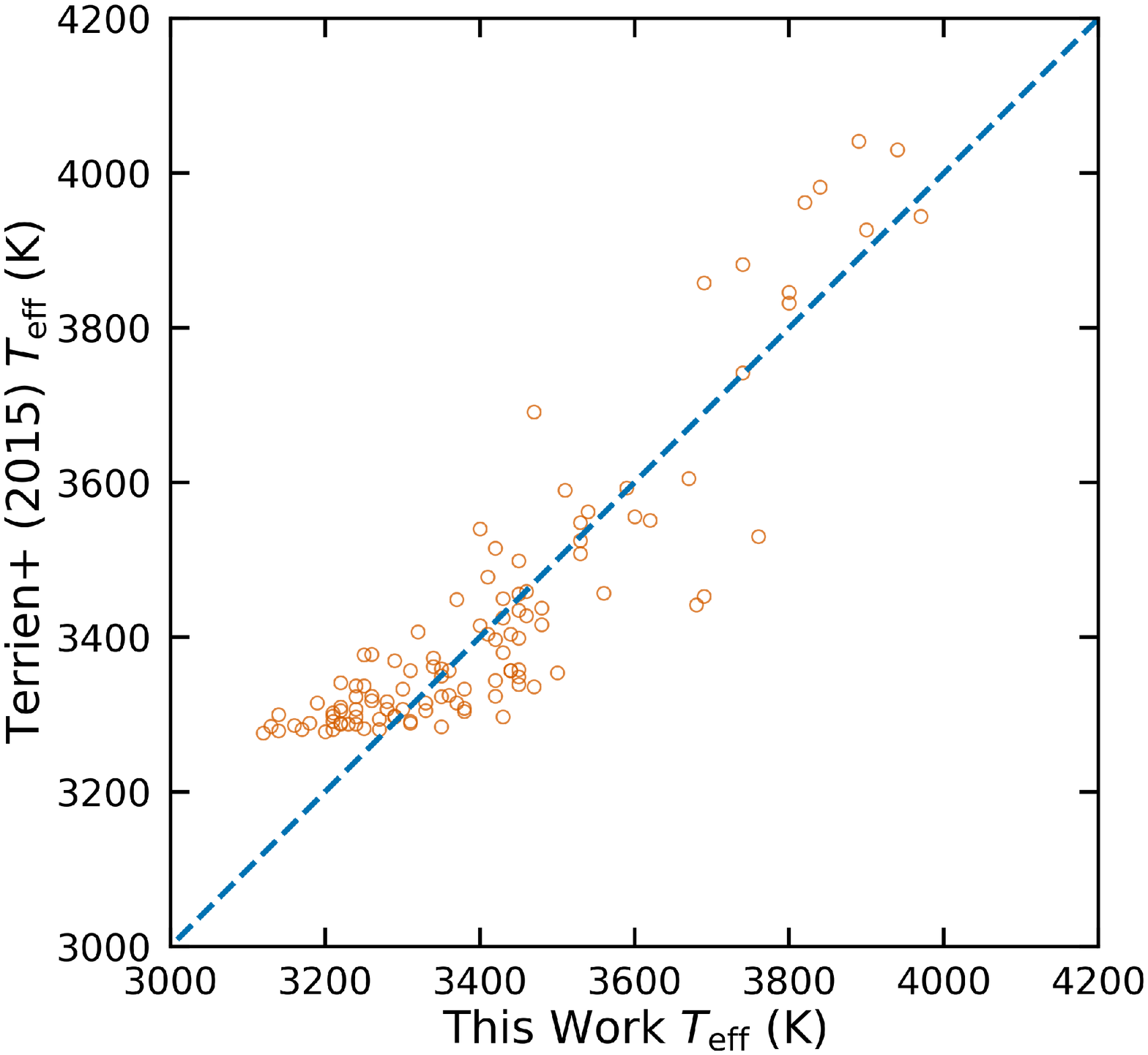}{0.47\textwidth}{\ \ \ \ \ \ \ \ \ \ (b)}}
\caption{(a) Comparison of infrared spectral types from \citet{Terrien2015} to our optical spectral types from LAMOST spectra. The optical spectral types are on average a half spectral type earlier than those from infrared spectra. (b) Effective temperatures from infrared and optical spectra are similar, however, the infrared spectra temperature relationships used in \citet{Terrien2015} are only valid down to a range of 3300\,K, which explains the sharp cut-off in the plot.}\label{fig:6}
\end{figure}

\citet{Terrien2015} conducted a near-infrared spectroscopic survey of 886 nearby M dwarfs, from which they identified spectral types and measured temperatures and metallicities. From this list, we found a matching LAMOST spectrum for 108 targets that match our criteria above, which allows us to compare results from our methods. \citet{Terrien2015} identified spectral types using a spectroscopic $\mathrm{H}_{2}\mathrm{O}$--K2 index typing method first used by \citet{Rojas-Ayala2012} and updated by \citet{Newton2014}. Our spectral types are on average a spectral type earlier than \citet{Terrien2015}, which is illustrated in Figure~\ref{fig:6}a. For consistency with the spectral typing of earlier-type stars, we recommend using spectral types based on optical spectra rather than infrared spectra when possible. Effective temperatures in \citet{Terrien2015} were measured using $K_s$ band index calibrations from \citet{Mann2013b}, which are valid in the range $3300\,\mathrm{K} < T_{\mathrm{eff}} < 4800\,\mathrm{K}$. We compared our derived temperatures in Figure~\ref{fig:6}b, which shows the sharp temperature cutoff in the \citet{Terrien2015} data at 3300\,K. Our temperatures are on average 20\,K less than those of \citet{Terrien2015}. Due to the similarity between temperature scales, we adopt the RMS scatter of 93\,K for our $T_{\mathrm{eff}}$ uncertainties.

\begin{figure*}[ht!]
\gridline{\fig{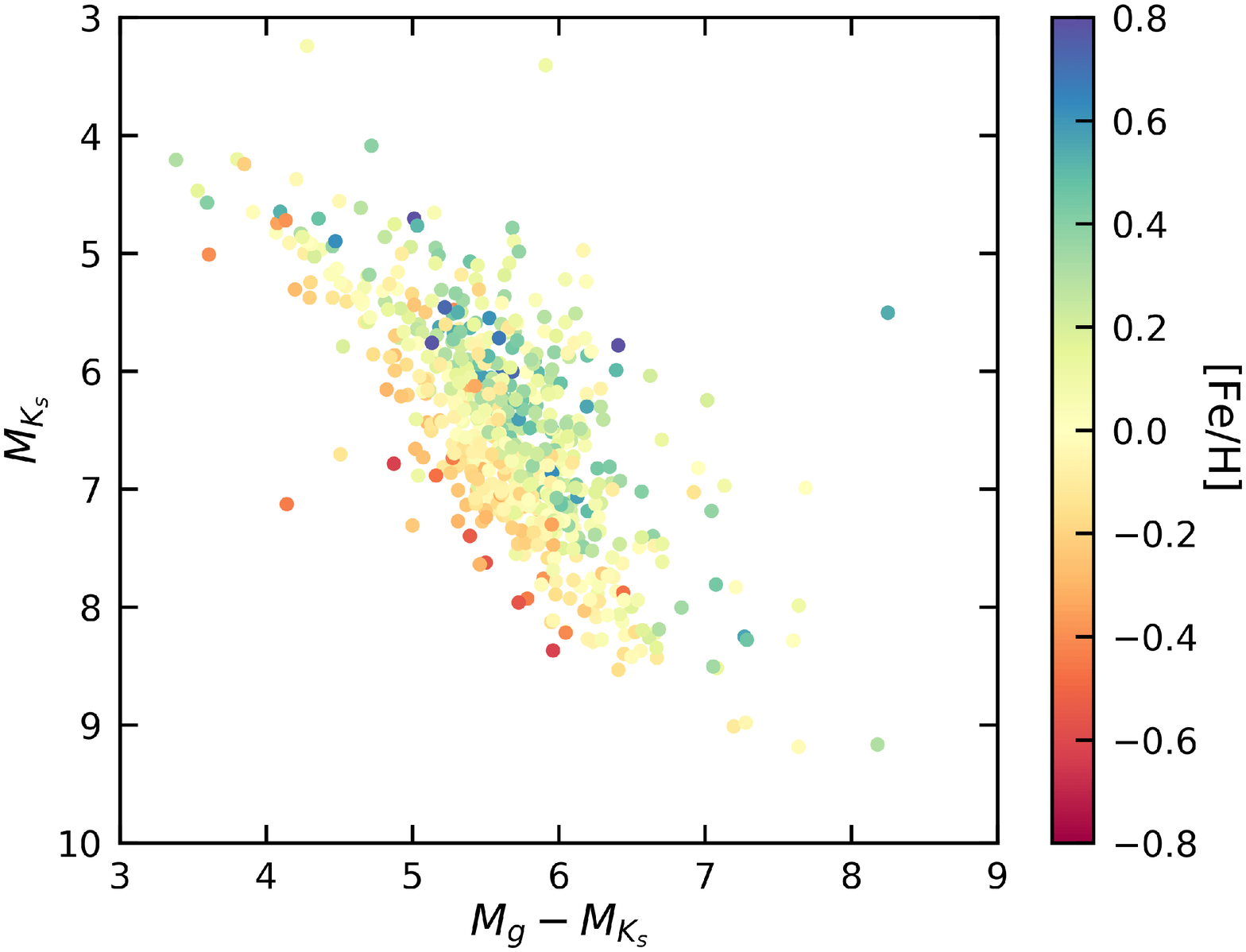}{0.48\textwidth}{(a)}
          \fig{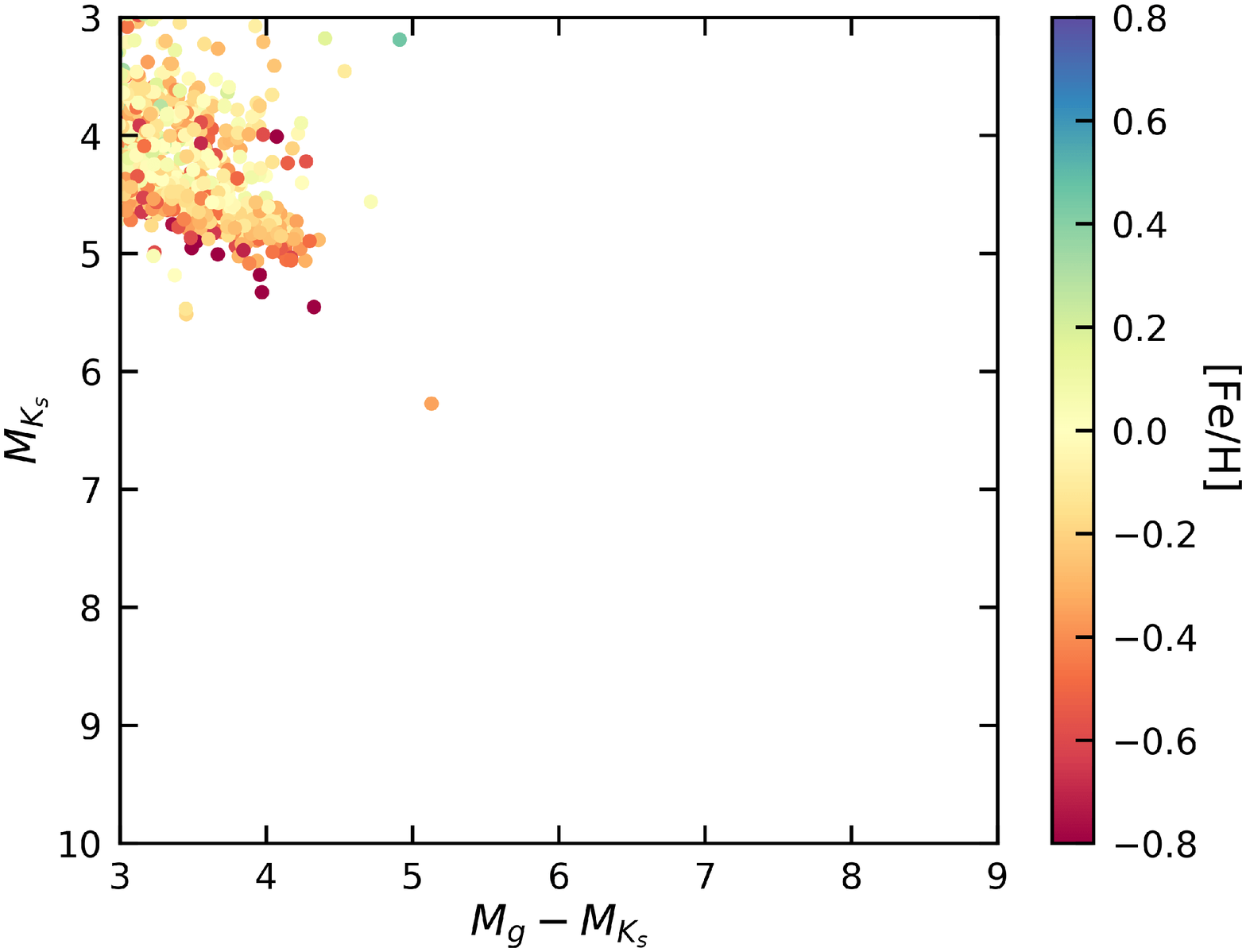}{0.48\textwidth}{(b)}}
\gridline{\fig{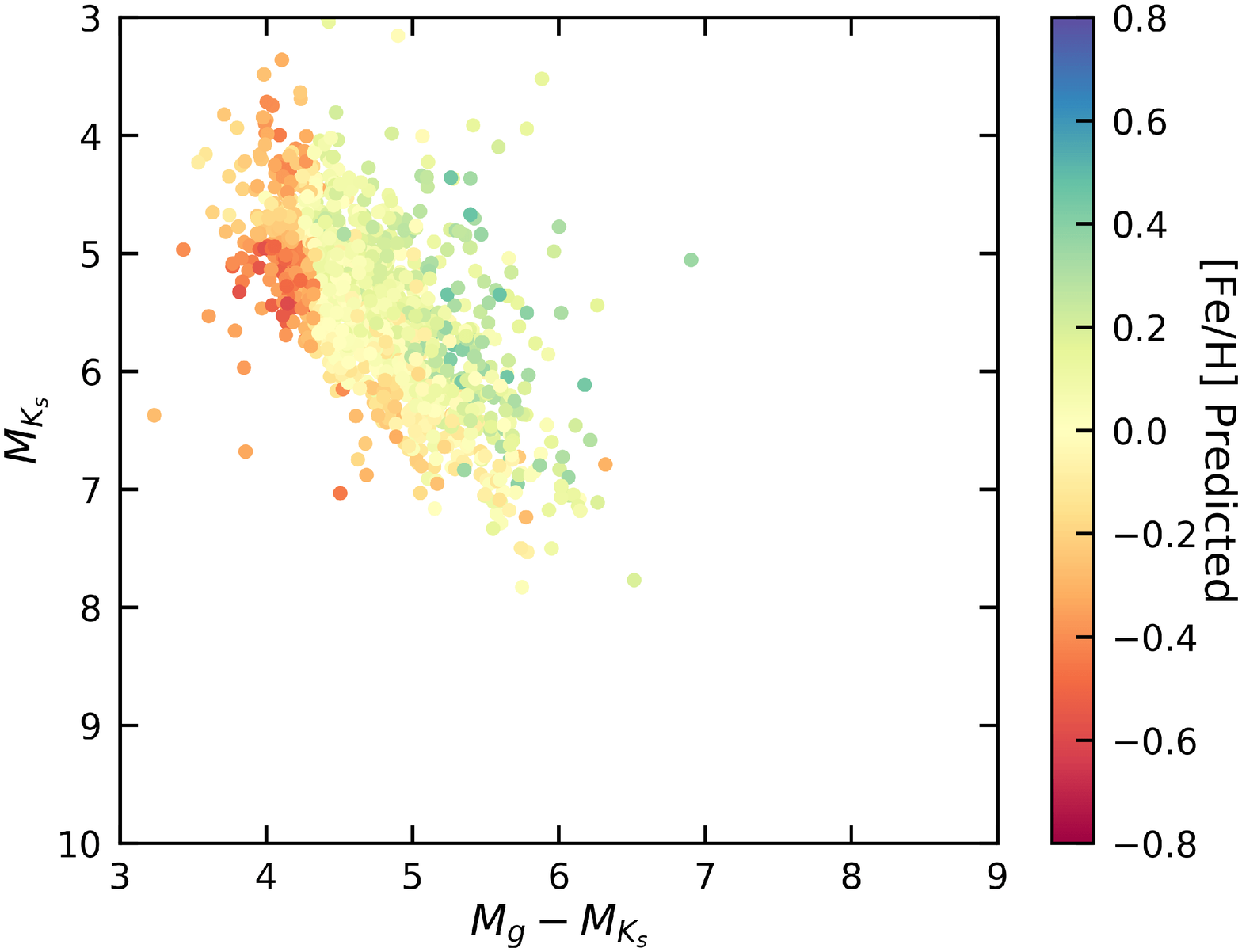}{0.48\textwidth}{(c)}
          \fig{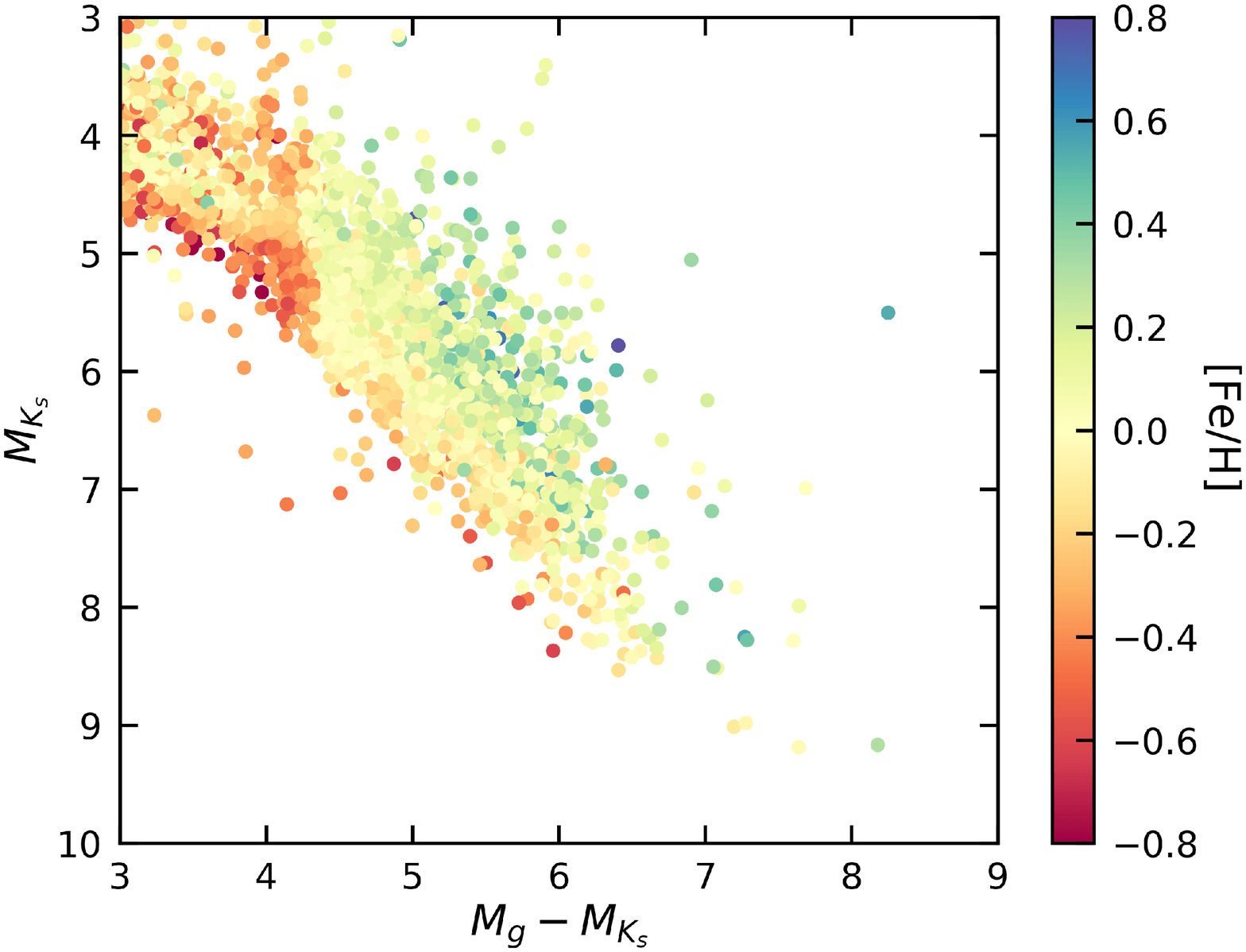}{0.48\textwidth}{(d)}}
\caption{Color-magnitude diagrams colored by [Fe/H] for the \citet{Terrien2015} M dwarf sample (a) and our LAMOST FGK targets (b). The results for the LAMOST M dwarf [Fe/H] classification are shown in (c), and these results are combined with (a) and (b) in panel (d).}\label{fig:7}
\end{figure*}

\subsection{Metallicity}\label{sec:feh}

The myriad molecular lines at optical wavelengths hinder the measurement of metallicity from moderate resolution optical spectra. Metallicity for M dwarfs can be directly measured if they have a wide-separation F, G, or K dwarf primary companion, assuming the stars formed at the same time from the same molecular cloud \citep{Bonfils2005}. These stars allow the calibration of absolute photometric \citep[e.g.,][]{Bonfils2005,Johnson2009,Schlaufman2010,Neves2012} and moderate resolution spectroscopic \citep[e.g.,][]{Rojas-Ayala2010,Terrien2012,Rojas-Ayala2012,Mann2013a,Newton2014,Mann2014} methods. From moderate resolution optical spectra, the $\zeta$ parameter, computed from TiO and CaH spectroscopic indices, has shown a weak correlation with metallicity \citep[e.g.,][]{Woolf2009,Mann2013a}, and the LAMOST pipeline provides measurements of $\zeta$ for M dwarfs. \citet{Mann2013a} compared different methods for computing M dwarf metallicities, and found that the highest quality calibrations come from $K$ band features from moderate resolution infrared spectra.

\begin{figure}[ht!]
    \gridline{\fig{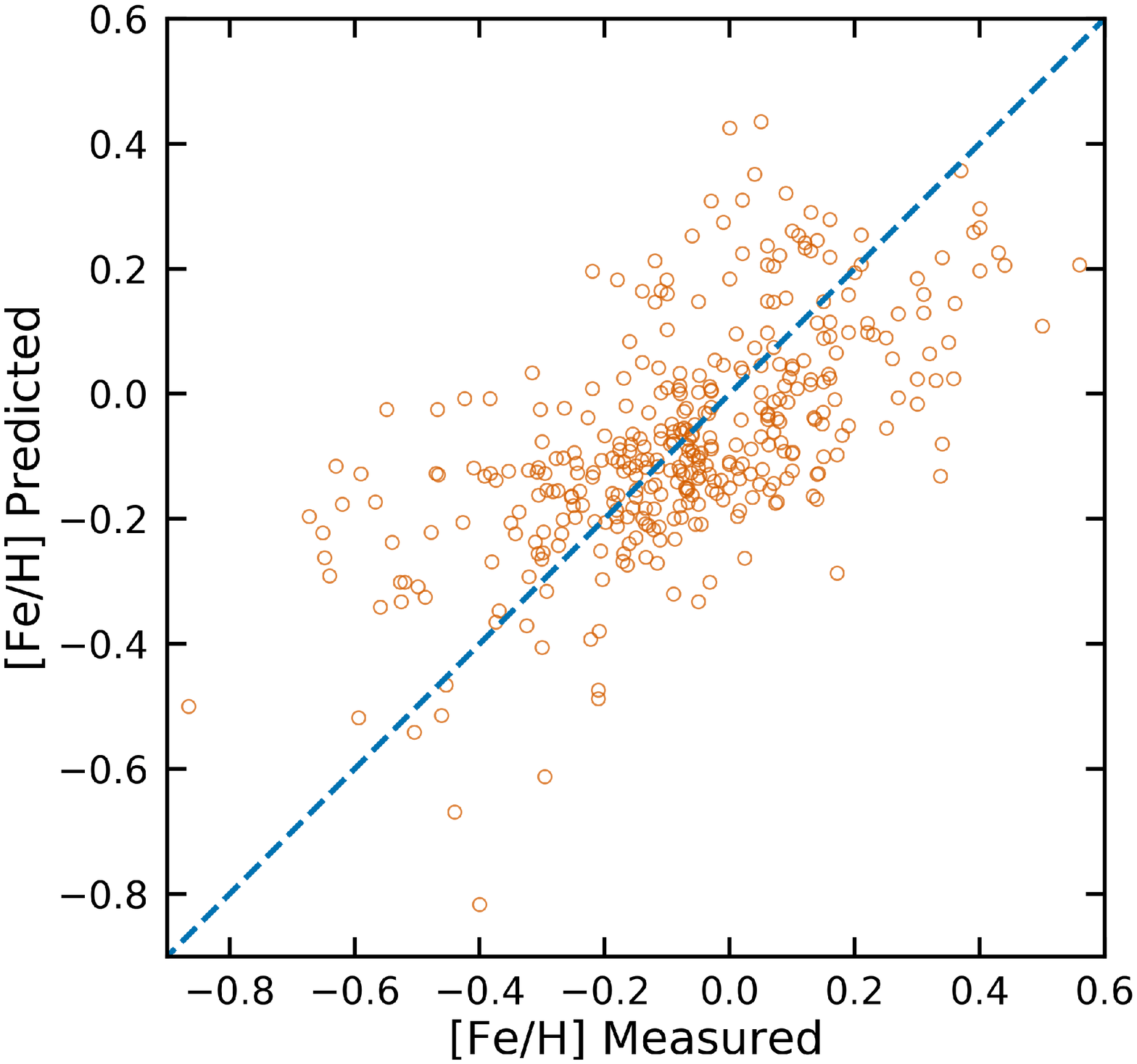}{0.46\textwidth}{\ \ \ \ \ \ (a)}}
    \gridline{\fig{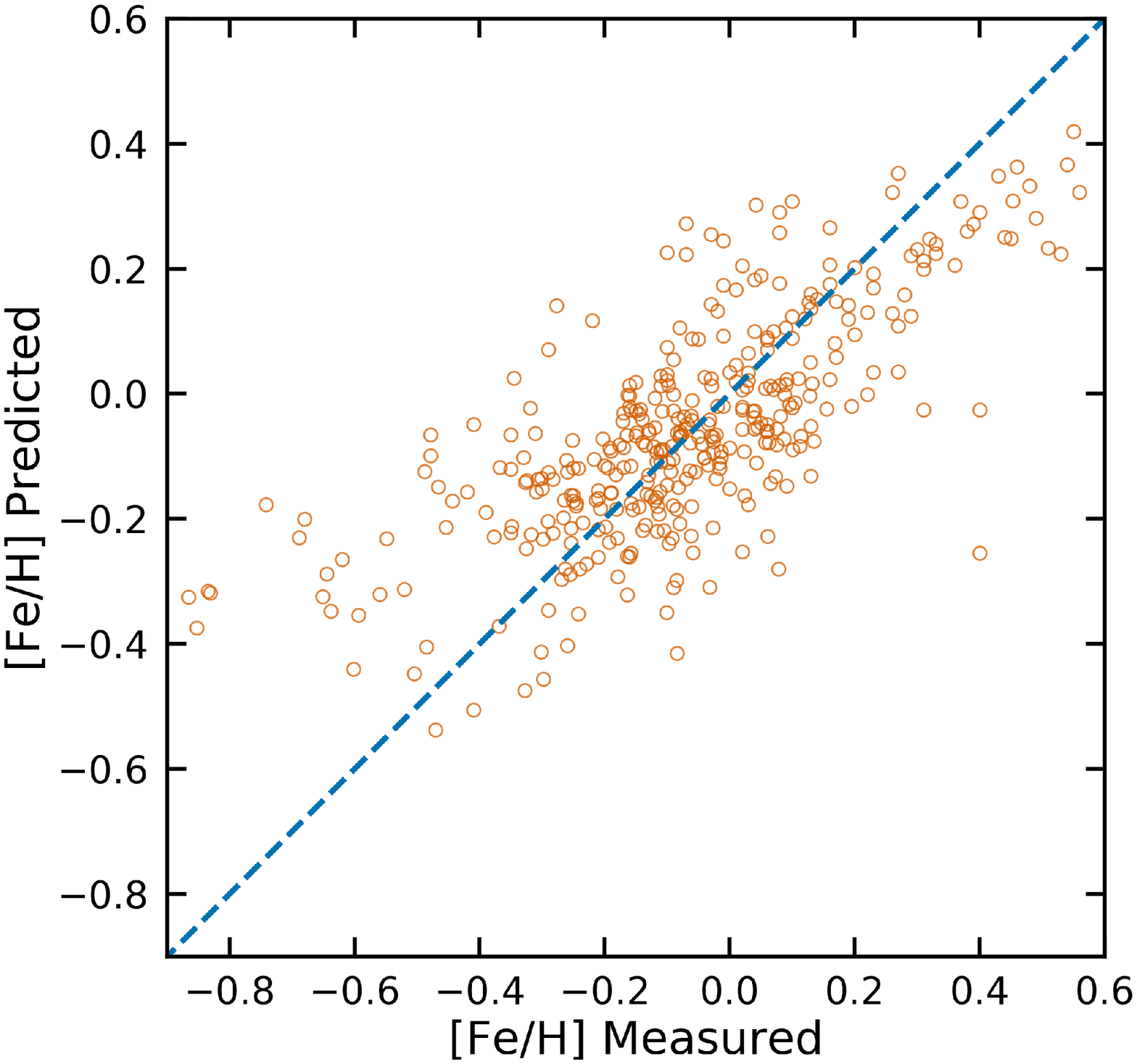}{0.46\textwidth}{\ \ \ \ \ \ (b)}}
    \caption{(a) Comparison of predicted versus measured [Fe/H] from our random forest regression using only $M_{K_s}$ and $M_g-M_{K_s}$ and (b) using $M_g-M_r$, $M_r-M_i$, $M_i-M_J$, $M_J-M_H$, and $M_H-M_{K_s}$, which yields a tighter correlation. The 1:1 lines are plotted for reference.}\label{fig:8}
\end{figure}

We initially tried to use the spectral indices and $\zeta$ measurements provided by the LAMOST pipeline to determine metallicity on the set of 108 stars with both a LAMOST spectrum and a $K$-band metallicity measurement from \citet{Terrien2015}, but we could not find any strong correlations. Instead, we calibrated a photometric metallicity relationship using 636 M dwarfs from \citet{Terrien2015} that have metallicity measurements, \gaia\ parallaxes, and $g$, $r$, $i$, $J$, $H$, and $K_s$-band photometry. We first computed absolute magnitudes for these targets. In Figure~\ref{fig:7}a, we plot $M_{K_s}$ versus $M_g-M_{K_s}$, with color indicating measured [Fe/H]. In this color space, there appears to be a metallicity gradient for M dwarfs, with larger $M_g-M_{K_s}$ colors generally indicating higher metallicity for the same $M_{K_s}$ magnitude. Due to the paucity of \citet{Terrien2015} targets with $M_g-M_{K_s} < 5$, we also included 1,483 of our LAMOST AFGK targets with measured metallicities, $M_{K_s} > 3$, and $M_g-M_{K_s} > 3$ (Figure~\ref{fig:7}b). We trained a random forest regressor \citep[\textsf{scikit-learn};][]{Pedregosa2011} with 1,000 trees on $M_{K_s}$ and $M_g-M_{K_s}$ for a random subset of 75\% of the 2,119 targets with measured metallicities. We used the remaining 25\% of targets to determine how well the regressor performed. Figure~\ref{fig:8}a compares the measured [Fe/H] to the predicted [Fe/H] from the random forest regressor. The median RMS scatter from 1,000 different random forest regressions using only $M_{K_s}$ and $M_g-M_{K_s}$ is 0.19. When we also included $M_g-M_r$, $M_r-M_i$, $M_i-M_J$, $M_J-M_H$, and $M_H-M_{K_s}$ as input parameters in the random forest regression, the median RMS scatter reduced to 0.17 (Figure~\ref{fig:8}b), which we took as the uncertainty in our M star [Fe/H] measurements. We plot the results from the [Fe/H] regression in Figure~\ref{fig:7}c and d.

\newpage
\subsection{Radius, Mass, and Surface Gravity}\label{sec:radmassgrav}

\begin{figure}[ht]
    \centering
    \includegraphics[width=0.48\textwidth]{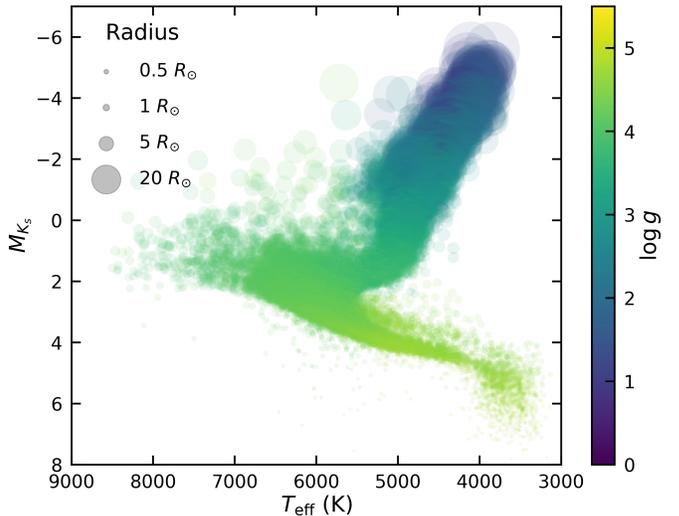}
    \caption{HR diagram LAMOST targets. Colors indicate surface gravity, and the size of the points represent stellar radius.} \label{fig:9}  
		\vspace{-0.5em}
\end{figure}

\citet{Mann2015} and \citet{Mann2019} derived empirical $M_{K_s}$--$R_{\star}$ and $M_{K_s}$--$M_{\star}$ relationships for M dwarfs to a precision below 3\%. We used these relationships to compute radii and masses of our M dwarfs. We added the model uncertainties from \citet{Mann2015} and \citet{Mann2019} in quadrature to our calculated Monte Carlo uncertainties, yielding average radius and mass uncertainties of 3.1\% and 6.6\%, respectively. From mass and radius, we calculated surface gravity for these stars using $\log\,g = \log(GM_{\star}/R_{\star}^2)$. We list all spectroscopically derived stellar parameters for AFGK and M stars in Table~\ref{tab:1} and show a Hertzsprung-Russell (HR) diagram of all LAMOST targets in Figure~\ref{fig:9}.

\section{Photometric Classifcation}\label{sec:photclass}

Using the 26,838 \kt\ targets classified from LAMOST spectra and \gaia\ parallaxes, we then classified stars with only photometry and \gaia\ parallaxes. The first step was to compute absolute magnitudes and the following colors to use for classification: $M_g-M_{K_s}$, $M_g-M_r$, $M_r-M_i$, $M_i-M_J$, $M_J-M_H$, and $M_H-M_{K_s}$. We first restricted our sample to \kt\ stars with these colors within the range of the LAMOST targets. This is necessary because random forest classification and regression cannot extrapolate beyond the range of the training set. This removed 9,673 targets from our sample, leaving us with 195,250 non-spectroscopic targets, and a total sample of 222,088 targets. A majority of the targets that were removed are fainter than $Kp=18$ (Figure~\ref{fig:1}).

\begin{figure}[ht]
    \centering
    \includegraphics[width=0.46\textwidth]{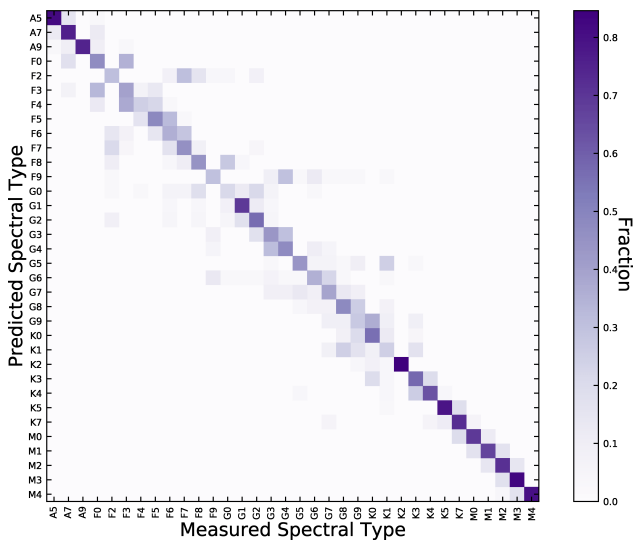}
    \caption{Measured versus predicted spectral types from our random forest classification, showing a reasonable prediction for most targets.} \label{fig:10}  
		\vspace{-0.5em}
\end{figure}

We began classification with spectral types. Table~\ref{tab:2} shows the number of targets with each spectral type in our LAMOST sample. Due to the relatively small numbers of A-type stars, we grouped A1-A6 stars into A5, and A8-A9 into A9 to increase the numbers in each respective bin for classification. In order to minimize bias due to different sample sizes, we randomly selected 100 stars from each spectral type to use for classification. For A5, A9, K2, and M4, we randomly sampled with replacement. In a similar manner to Section~\ref{sec:feh}, we used these aforementioned colors along with $M_{K_s}$ to train a random forest classifier \citep[\textsf{scikit-learn};][]{Pedregosa2011} with 1,000 trees on a random subset of 75\% of the spectroscopic target subsample. The remaining 25\% of the subsample were used to check the classifier performance. Figure~\ref{fig:10} shows the measured versus predicted spectral type from the testing set. A majority of the predicted classifications are along or near the diagonal, indicating the classifier does a reasonable job at predicting spectral type. We used the trained classifier on all the photometric targets to yield spectral types. The assigned spectral types from photometry should be adequate for large statistical studies of \kt\ targets, but we caution their use for individual targets, and strongly encourage obtaining a spectrum for accurate spectral typing.

\begin{figure}[ht!]
\gridline{\fig{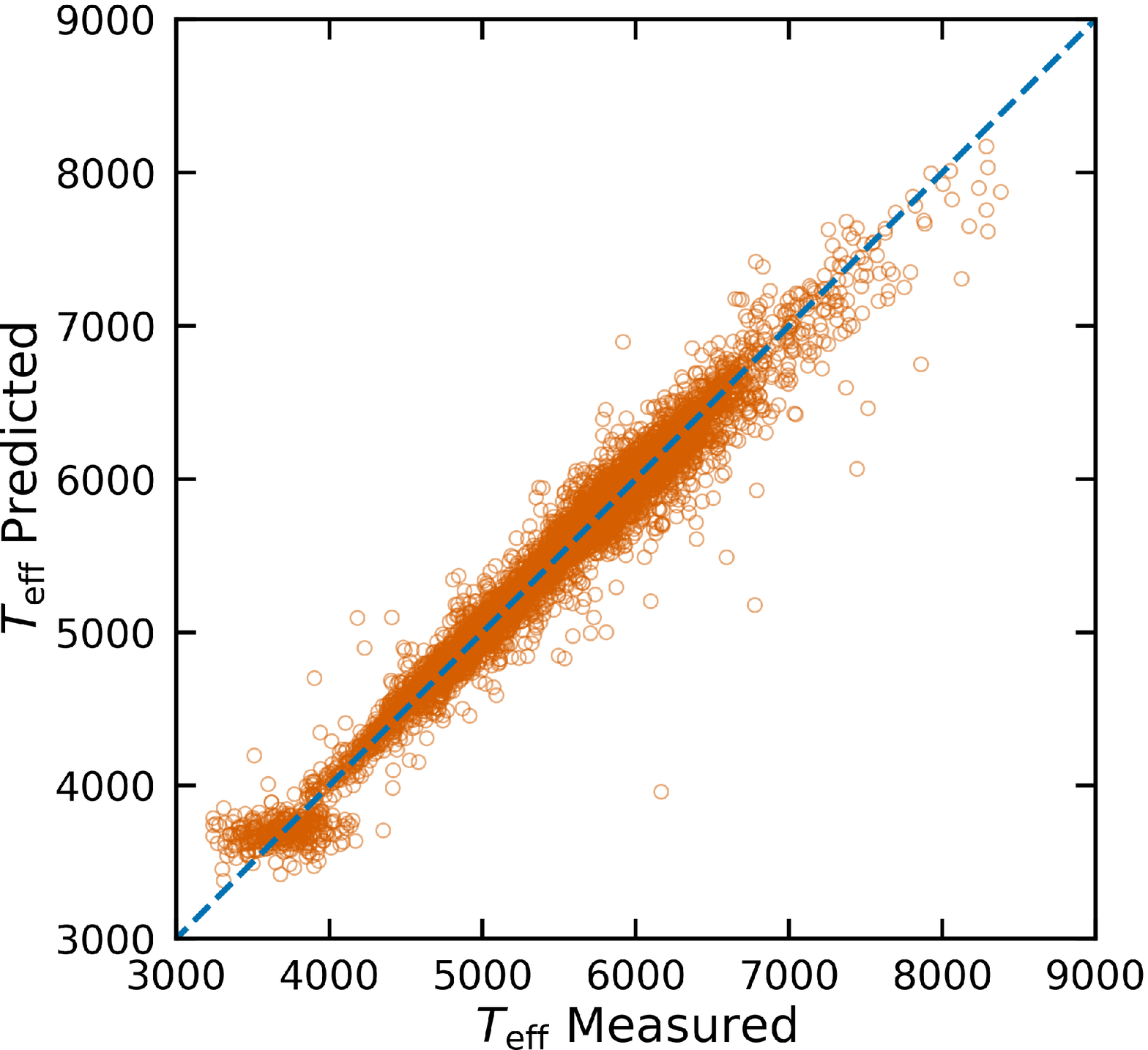}{0.345\textwidth}{\ \ \ \ \ \ (a)}}\vspace{-0.25em}
\gridline{\fig{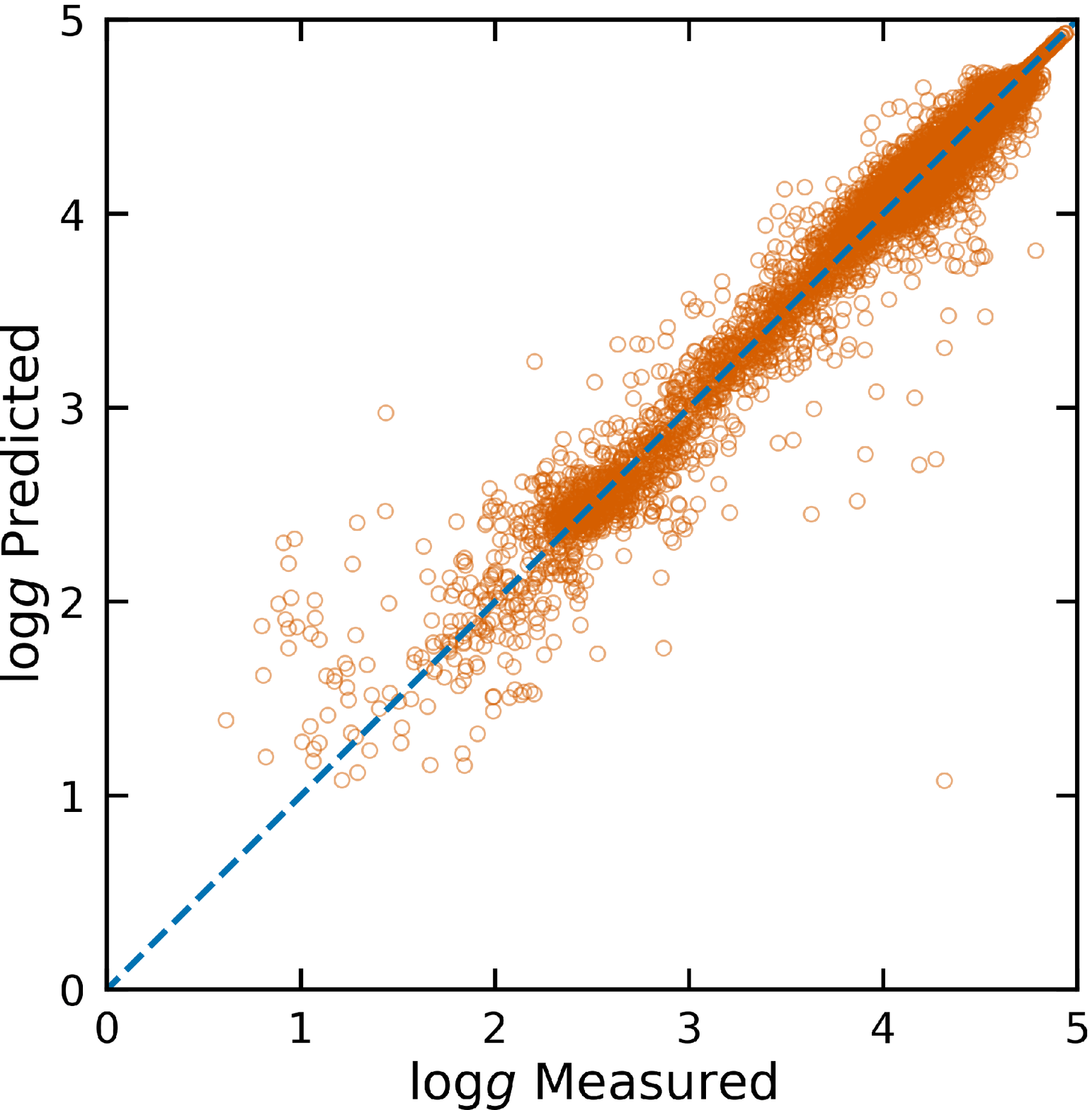}{0.3143\textwidth}{\ \ \ \ \ \ (b)}}\vspace{-0.25em}
\gridline{\fig{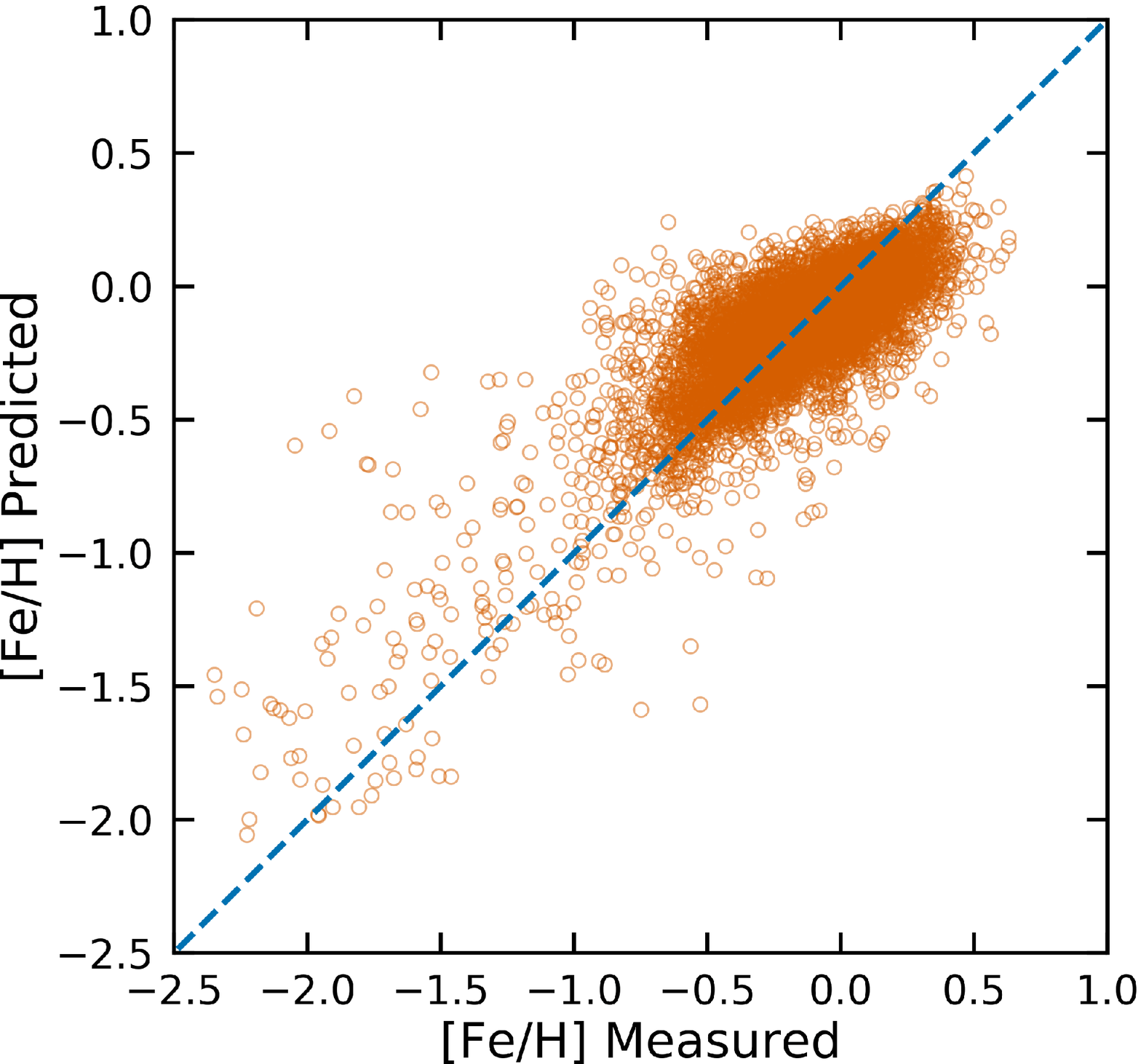}{0.338\textwidth}{\ \ \ \ \ \ (c)}}\vspace{-0.25em}
\caption{Comparison of predicted versus measured $T_{\mathrm{eff}}$ (a), $\log\,g$ (b), and [Fe/H] (c) from our random forest regression using $M_{K_s}$, $M_g-M_{K_s}$, $M_g-M_r$, $M_r-M_i$, $M_i-M_J$, $M_J-M_H$, and $M_H-M_{K_s}$. The 1:1 lines are plotted for reference. There are tight correlations for $T_{\mathrm{eff}}$ and $\log\,g$, and a positive correlation for [Fe/H].}\label{fig:11}
\end{figure}

For effective temperature, surface gravity, and metallicity, we followed the same procedure outlined in Section~\ref{sec:feh}, training a random forest regressor on $M_{K_s}$, $M_g-M_{K_s}$, $M_g-M_r$, $M_r-M_i$, $M_i-M_J$, $M_J-M_H$, and $M_H-M_{K_s}$ for our targets with spectroscopic $T_{\mathrm{eff}}$, $\log\,g$, and [Fe/H] measurements. Figure~\ref{fig:11} shows the results from the testing set, with good fits for $T_{\mathrm{eff}}$ and $\log\,g$, and a positive correlation for [Fe/H]. We adopted the RMS scatter as the uncertainties for photometrically classified targets, which are 138\,K, 0.15 dex, and 0.20 dex for $T_{\mathrm{eff}}$, $\log\,g$, and [Fe/H], respectively. Stellar radii and masses were then computed using the same procedures outlined in Section~\ref{sec:afgk} for AFGK stars, and Section~\ref{sec:radmassgrav} for M stars. Average uncertainties on $R_{\star}$ and $M_{\star}$ for photometrically classified targets are 7\% and 38\%, respectively. We list the parameters for stars classified using photometry in Table~\ref{tab:1}.

\section{Discussion}\label{sec:discussion}
\subsection{Comparison to previous stellar measurements}\label{sec:previous}
\begin{figure*}[hp]
    \gridline{\fig{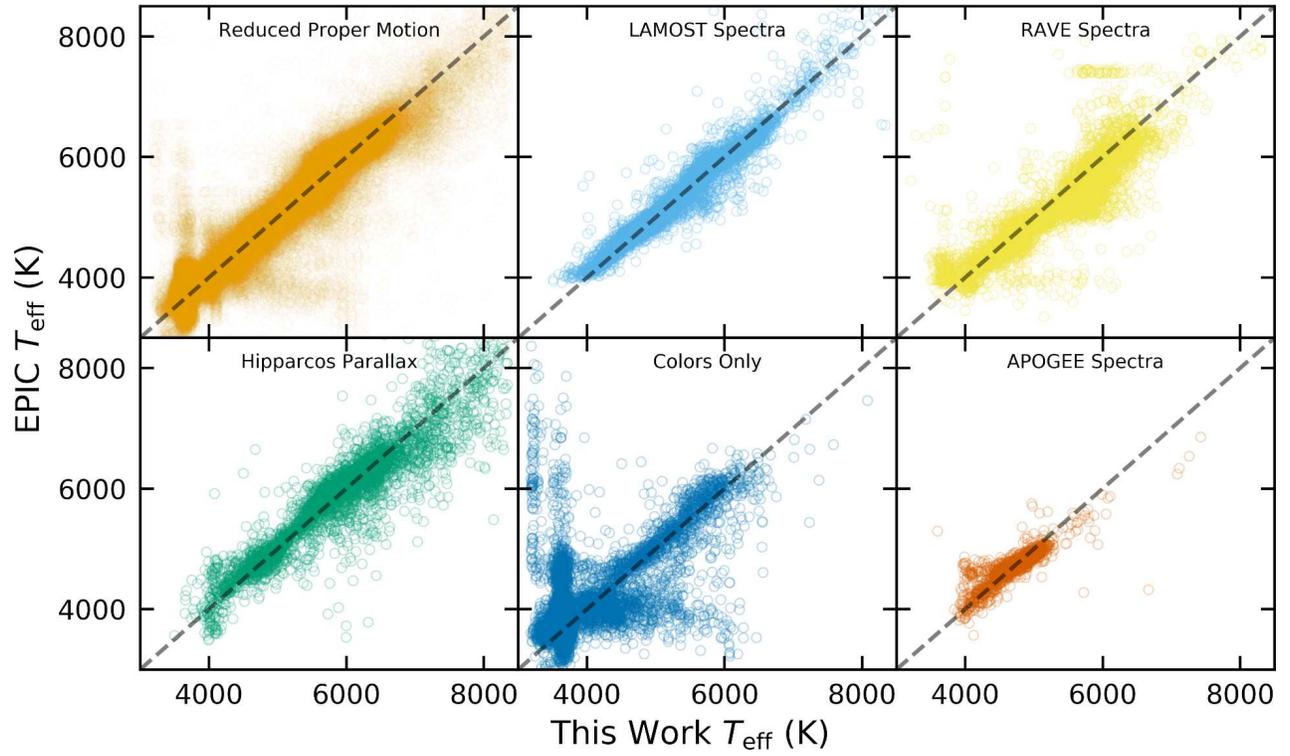}{0.93\textwidth}{(a)}}\vspace{-0.5em}
    \gridline{\fig{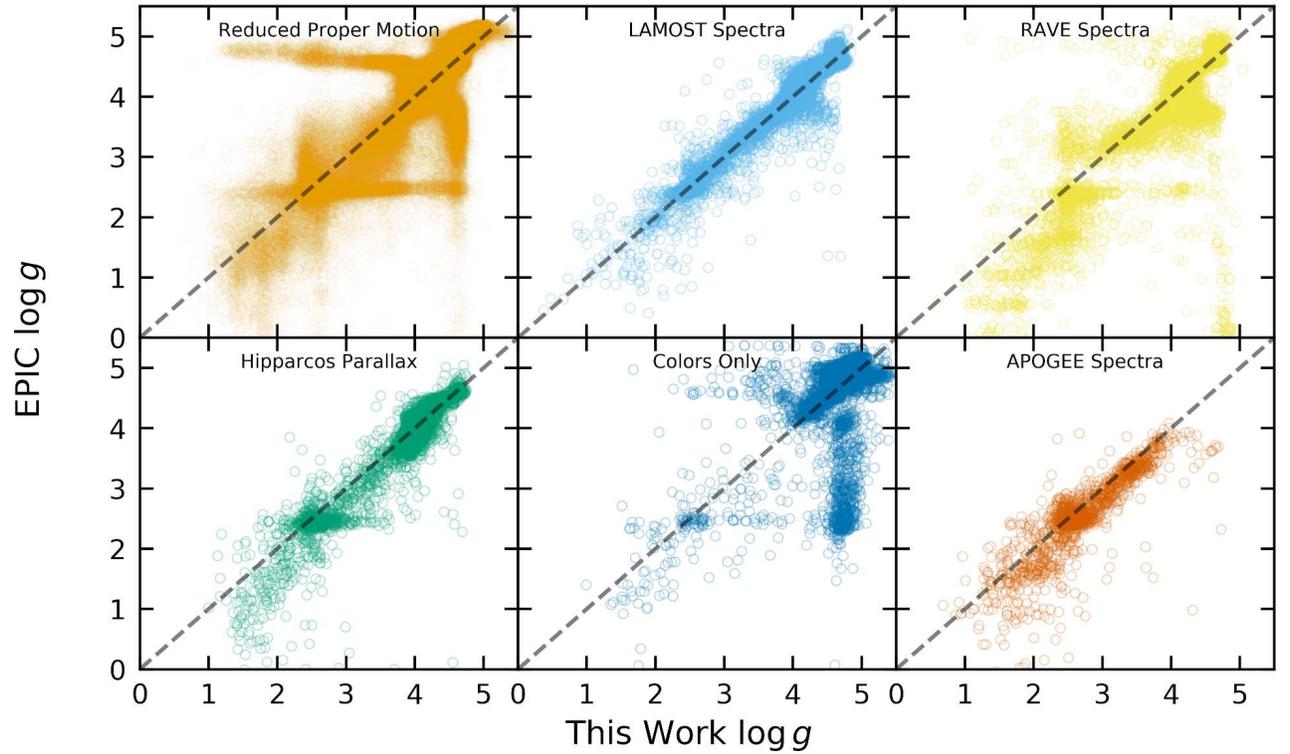}{0.93\textwidth}{(b)}}\vspace{-0.5em}
    \caption{Comparison of our temperature (a) and surface gravity (b) measurements to those from the EPIC. Each panel compares our measurements to the different methods used to derive the parameters in the EPIC, which allows us to elucidate any potential trends based on classification method.}\label{fig:12}
\end{figure*}

\begin{figure*}[hp]
\gridline{\fig{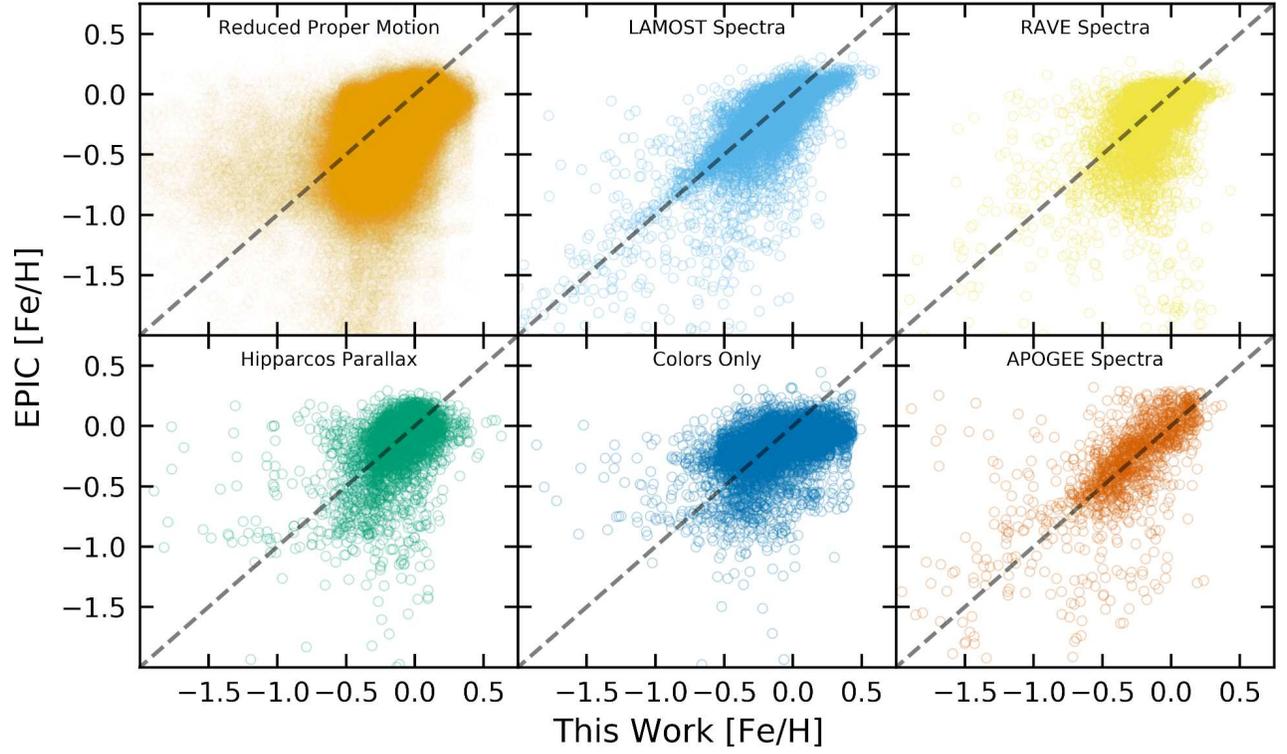}{0.93\textwidth}{(a)}}\vspace{-0.5em}
\gridline{\fig{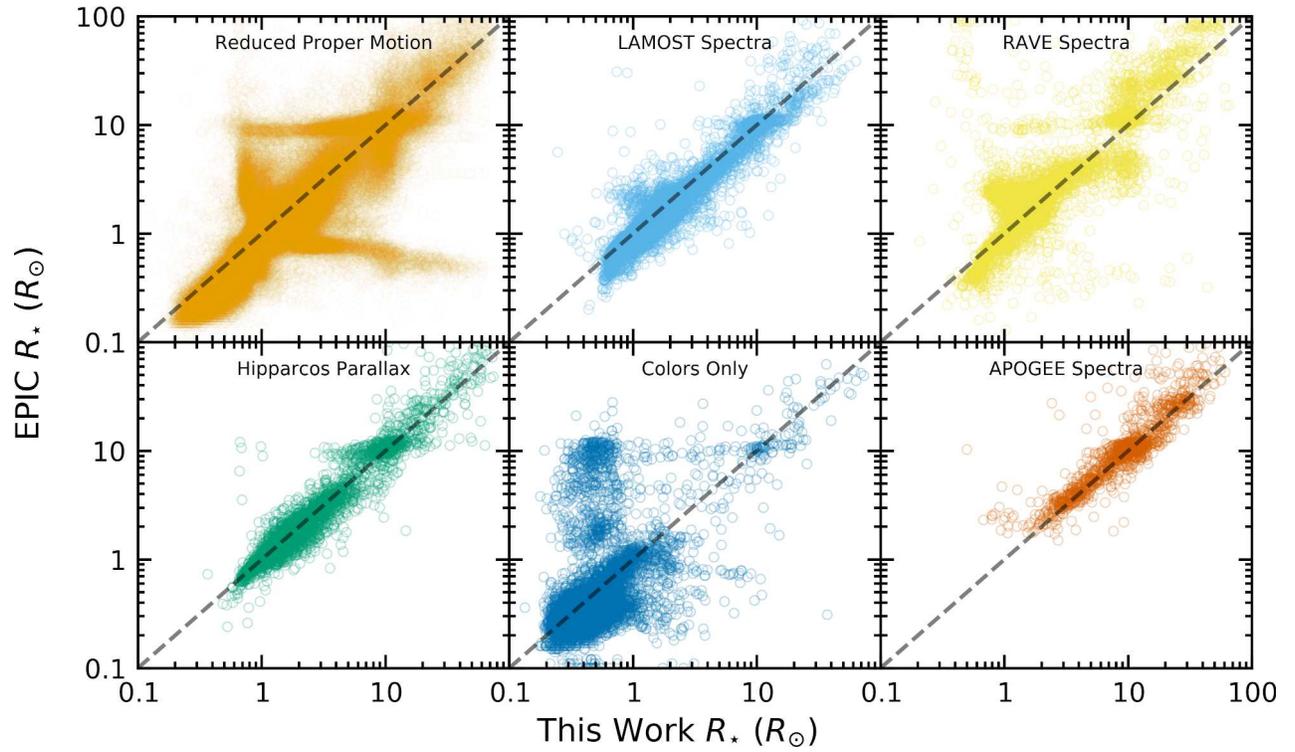}{0.93\textwidth}{(b)}}\vspace{-0.5em}
\caption{Same as Figure~\ref{fig:12}, but for metallicity (a) and stellar radii (b).}\label{fig:13}
\end{figure*}

\begin{figure*}[ht]
    \centering
    \includegraphics[width=0.93\textwidth]{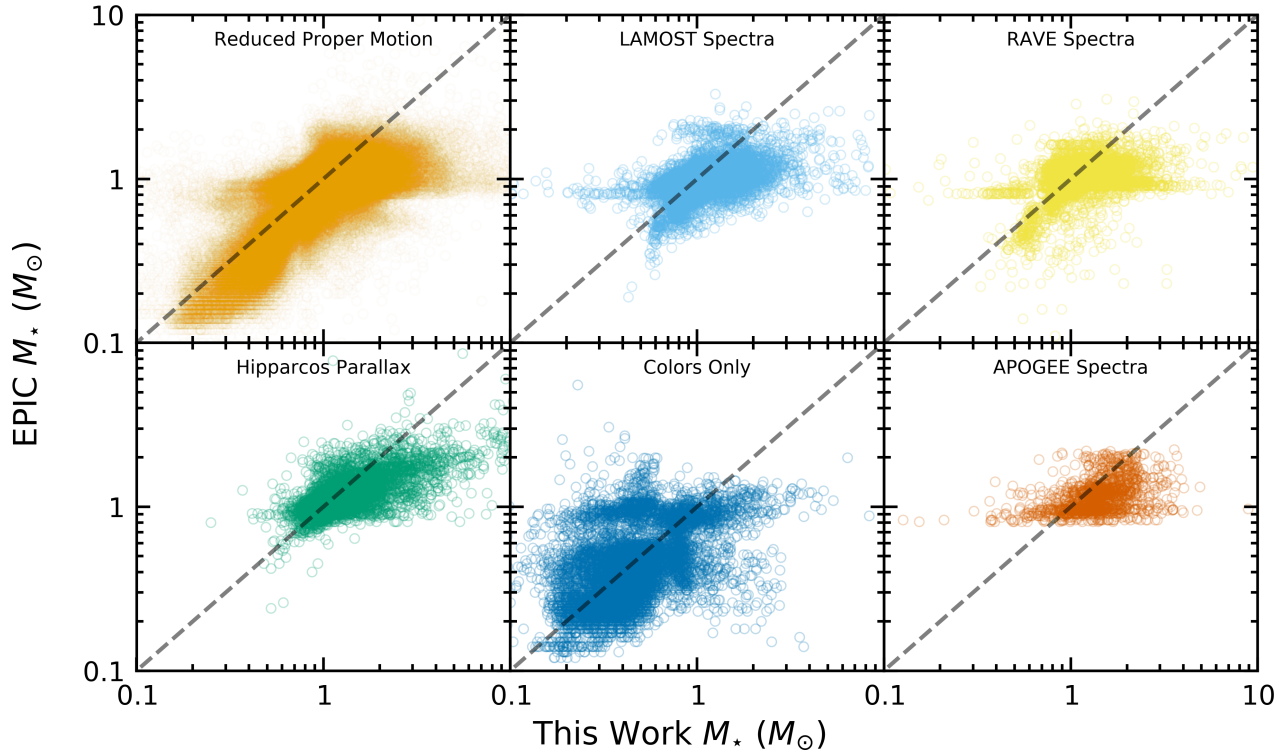}
    \caption{Same as Figure~\ref{fig:12}, but for stellar mass.} \label{fig:14}  
		\vspace{-0.5em}
\end{figure*}

The EPIC contains $T_{\mathrm{eff}}$, $\log\,g$, [Fe/H], $R_{\star}$, and $M_{\star}$ measurements for 192,598 of our targets, which allowed us to compare results. A significant fraction of the stellar properties for these targets in the EPIC were measured using reduced proper motions and colors (165,641), with LAMOST spectra accounting for 8,115 targets, RAVE spectra:\ 4,938 targets, APOGEE spectra:\ 1,413 targets, \emph{Hipparcos} parallax:\ 4,912 targets, and colors only:\ 7,579 targets. In Figures~\ref{fig:12}, \ref{fig:13}, and \ref{fig:14} we compare our $T_{\mathrm{eff}}$, $\log\,g$, [Fe/H], $R_{\star}$, and $M_{\star}$ measurements to those from the EPIC, delineating between the different EPIC classification inputs to see if there are any major trends depending on classification method. In general, our effective temperatures are similar regardless of classification method. For surface gravity there is much more structure, with a few preferential `arms' appearing where there are significant interchanges between dwarfs and giants. There is a positive correlation between the measurements of [Fe/H], but in general our measurements appear to be larger. In the $R_{\star}$ comparisons, the giant-dwarf interchange arms are again apparent in the reduced proper motion and colors only plots. There are positive correlations between the mass measurements, but our mass measurements are generally larger than EPIC values.

The parameters derived from LAMOST spectra measurements are unsurprisingly similar, with deviations from unity mostly caused by our measurements of M dwarf properties. It is worth noting that our LAMOST measurements are from DR5, whereas the EPIC values come from LAMOST DR1. LAMOST pipeline updates changed computed parameters, and a comparison between LAMOST DR5 and DR3 for the same targets showed a standard deviation of 83\ K, 0.13 dex, and 0.07 dex for $T_{\mathrm{eff}}$, $\log\,g$, and [Fe/H], respectively\footnote{\href{http://dr5.lamost.org/doc/release-note-v2}{http://dr5.lamost.org/doc/release-note-v2}}.

\begin{figure*}[ht!]
    \gridline{\fig{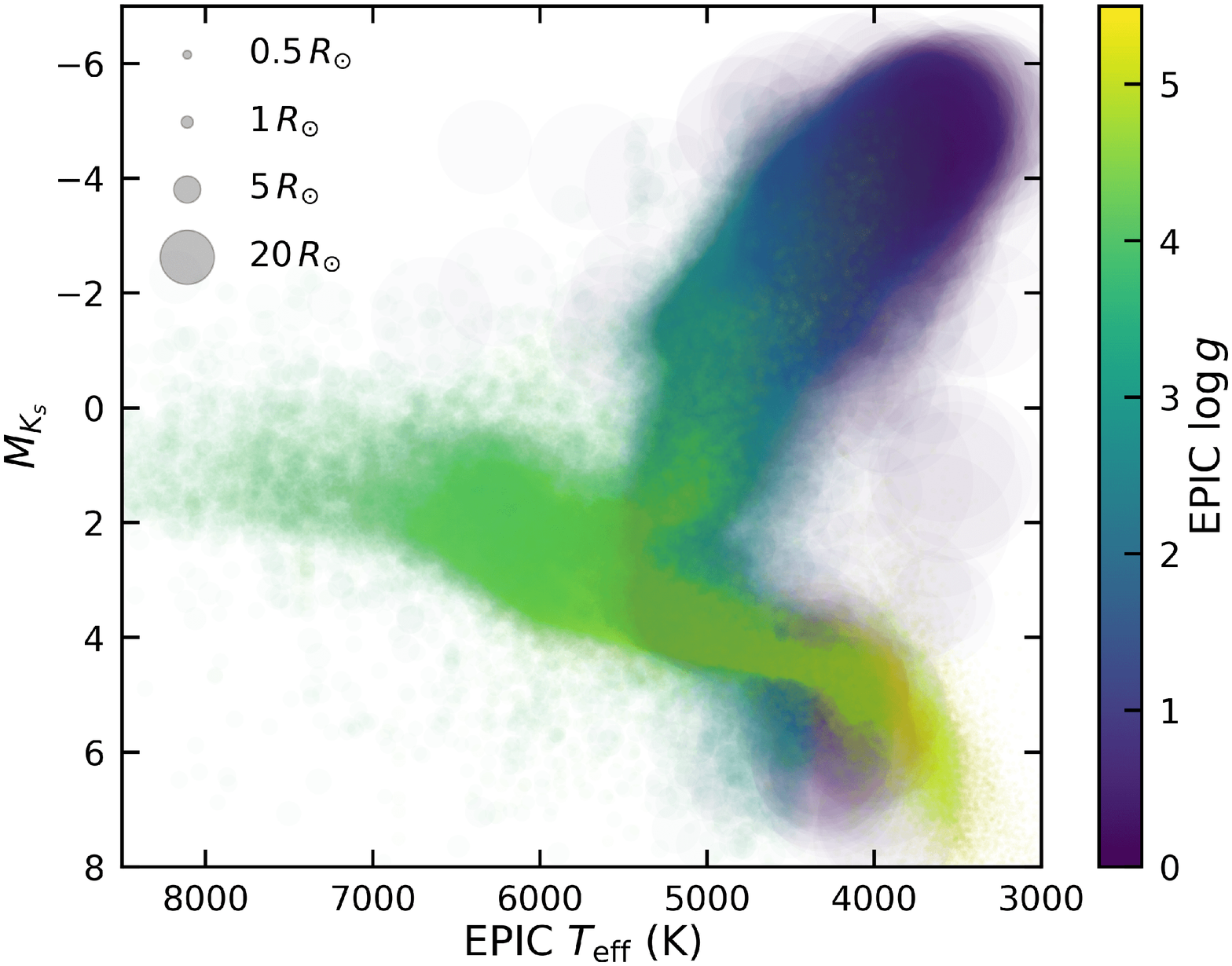}{0.48\textwidth}{(a)}
              \fig{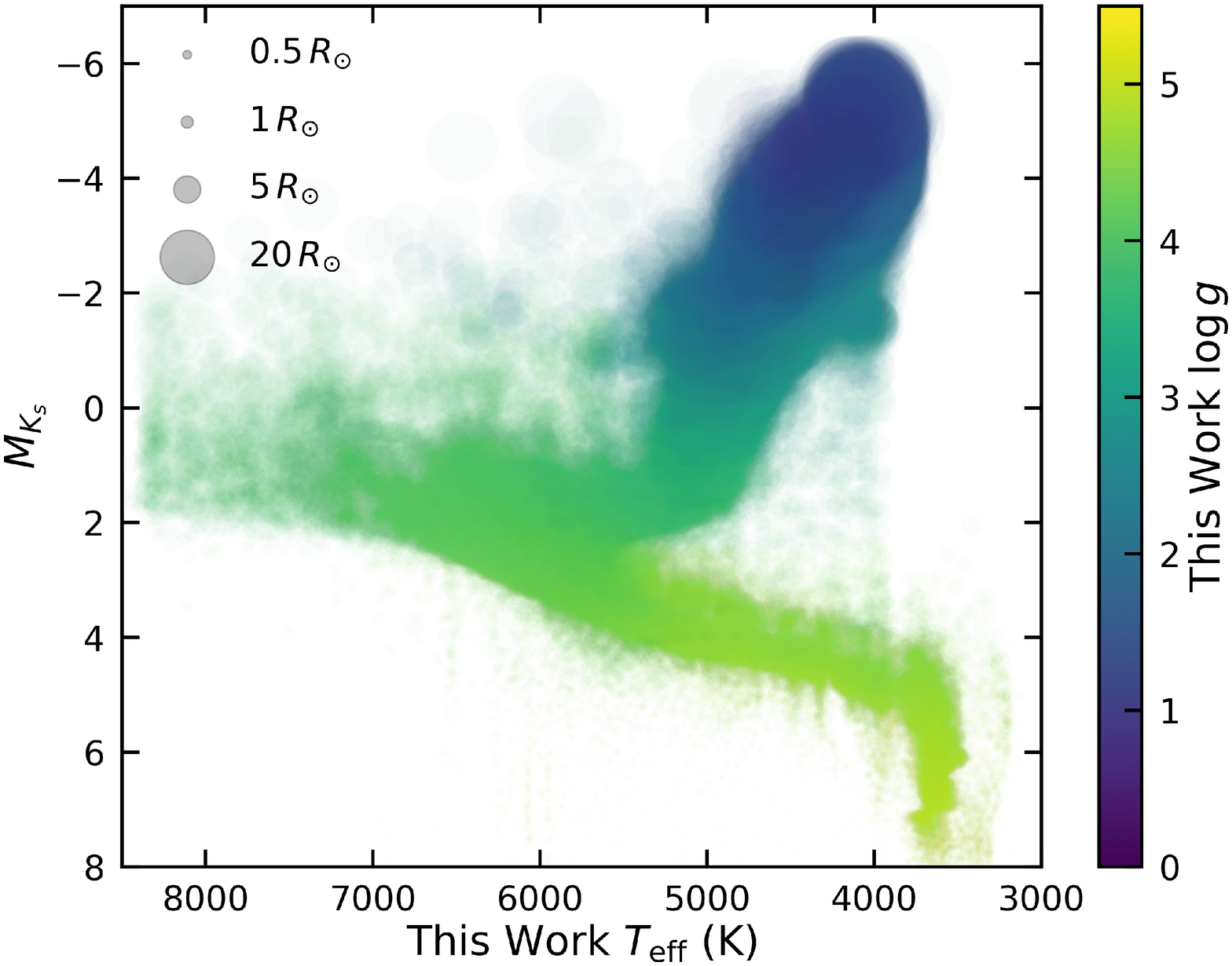}{0.48\textwidth}{(b)}}
    \caption{HR diagrams for (a) EPIC parameters and (b) our parameters. Colors indicate surface gravity, and the size of the points represent stellar radius. Several giant-dwarf interlopers are clearly visible in the EPIC HR diagram.}\label{fig:15}
\end{figure*}

Using our $M_{K_s}$ values, we compare HR diagrams for $T_{\mathrm{eff}}$ in the EPIC and our values in Figure~\ref{fig:15}, showing additional information from surface gravities and radii. The aforementioned giant--dwarf misclassifications are clearly visible in the EPIC HR diagram.

Since there were no M giants in our LAMOST sample, it is difficult to accurately classify these targets for \kt. M giants will have similar colors to M dwarfs, but very different luminosities. Table~\ref{tab:1} contains a few hundred low surface gravity targets ($1.2 \lesssim \log\,g \lesssim 3.9$) with an assigned M spectral type. Notably, these targets have temperatures higher than $\sim$4200\,K, likely due to the random forest regressor assigning temperatures of nearby K giants with similar $M_{K_s}$ magnitudes. We urge caution when using our catalog parameters for targets toward the tip of the giant branch, and recommend using surface gravity and absolute magnitudes to help differentiate between main sequence and evolved stars.

\begin{figure*}[hp]
    \gridline{\fig{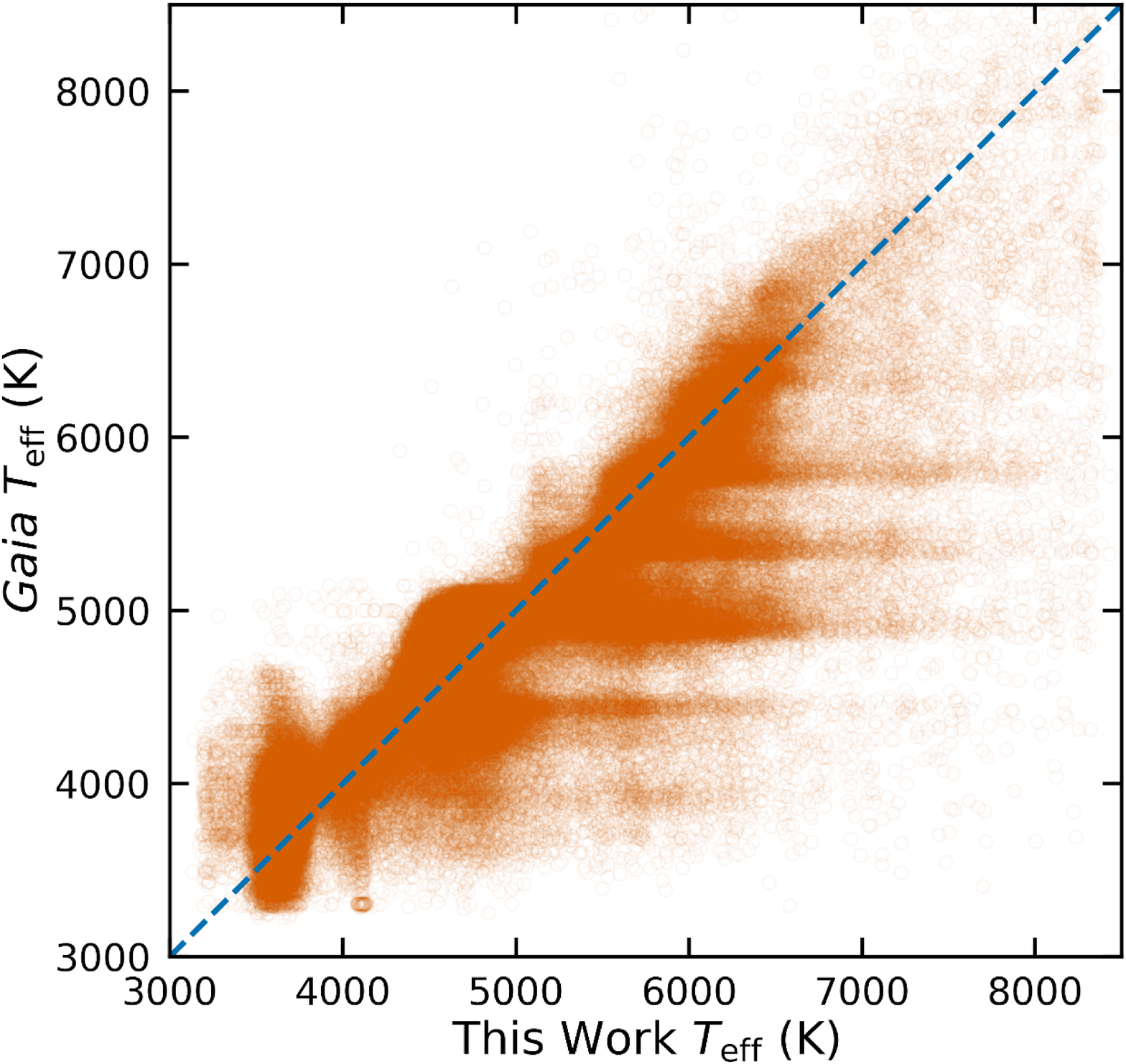}{0.48\textwidth}{\ \ \ \ \ \ (a)}
              \fig{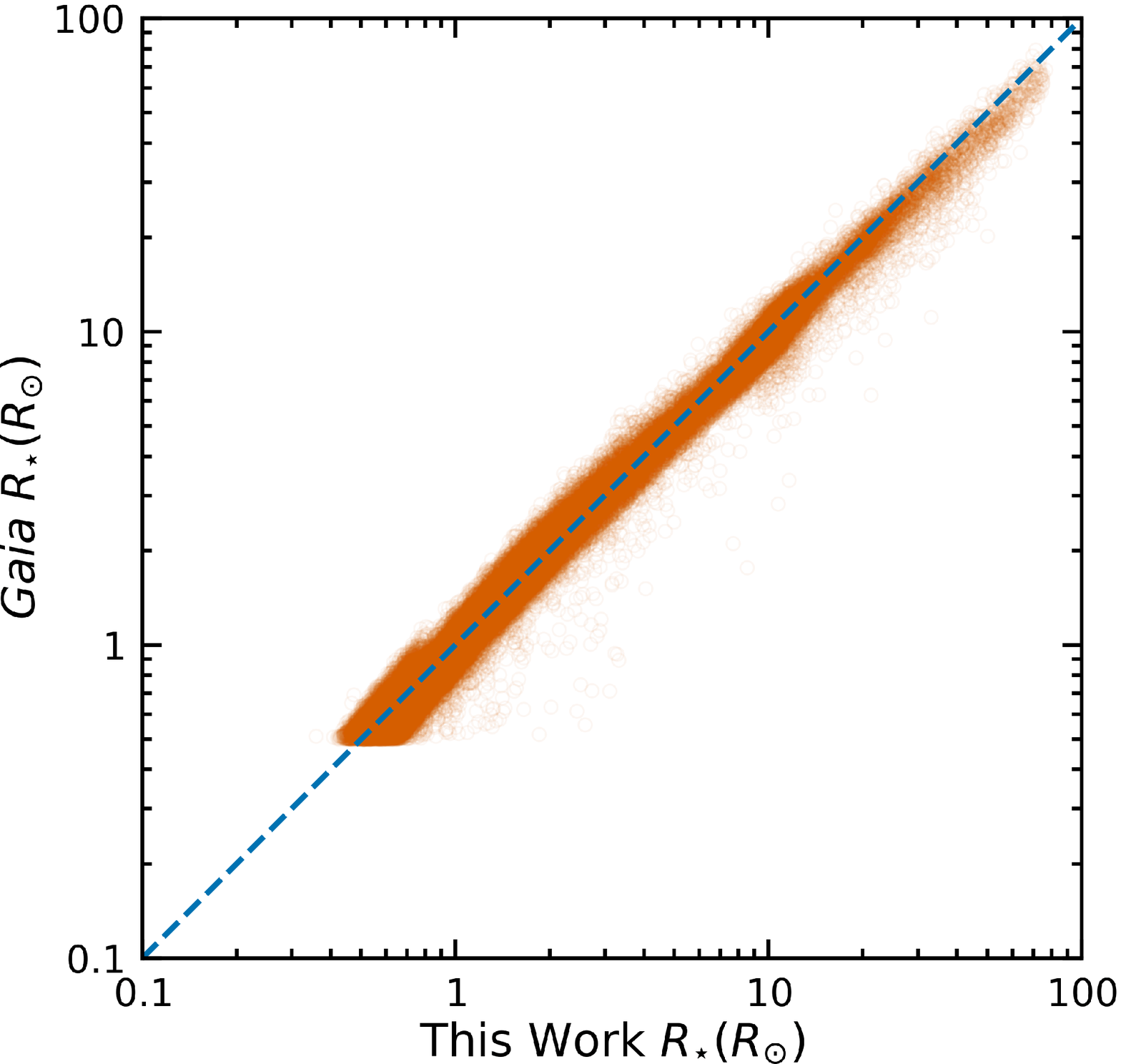}{0.48\textwidth}{\ \ \ \ \ \ (b)}}
    \caption{Comparison of our temperature (a) and stellar radius (b) measurements to those from \gaia. We caution readers to be careful when using \gaia\ effective temperatures. It is also notable that \gaia\ does not contain radius measurements for most M dwarfs smaller than $0.5\,R_{\odot}$.}\label{fig:16}
\end{figure*}

\gaia\ measured $T_{\mathrm{eff}}$ and $R_{\star}$ for 174,781 of our stars, which we compare in Figure~\ref{fig:16}. The \gaia\ temperatures were estimated using $G$, $G_{BP}$, and $G_{RP}$ colors using a random forest algorithm trained on stars with $T_{\mathrm{eff}}$ determined from spectra \citep{Andrae2018}. In general, our $T_{\mathrm{eff}}$ measurements are comparable to \gaia\ measurements, but there appear to be more preferential temperatures in the \gaia\ targets, likely caused by their input training set. Our stellar radii correlate well with those determined from \gaia\, which were measured in a similar manner to ours from the Stefan-Boltzmann law, using $M_G$ instead of $M_{K_s}$. Notably absent from \gaia\ measured radii are stars below $0.5\ R_{\odot}$.

\begin{figure*}[hp]
    \gridline{\fig{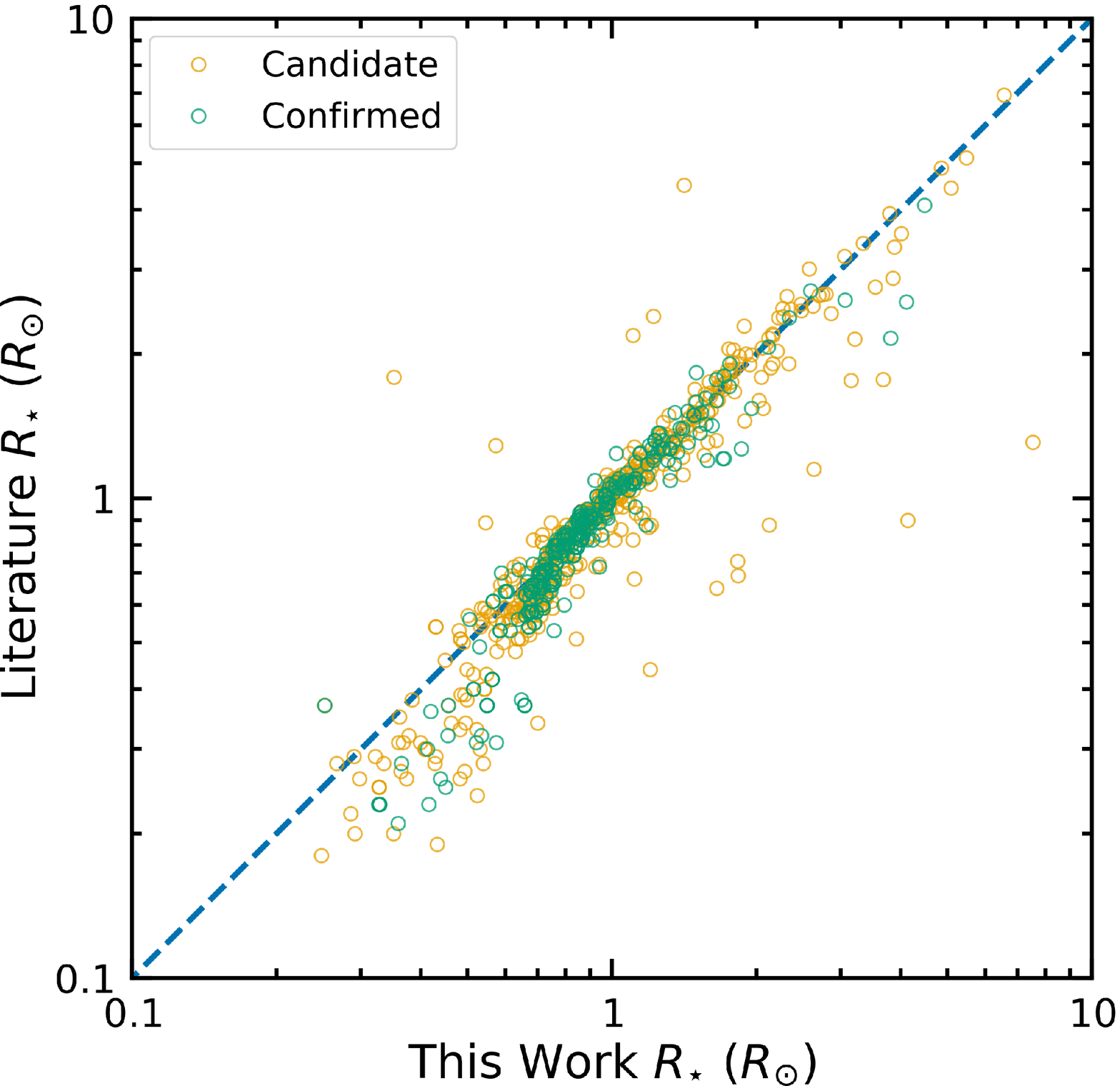}{0.45\textwidth}{\ \ \ \ \ \ (a)}
              \fig{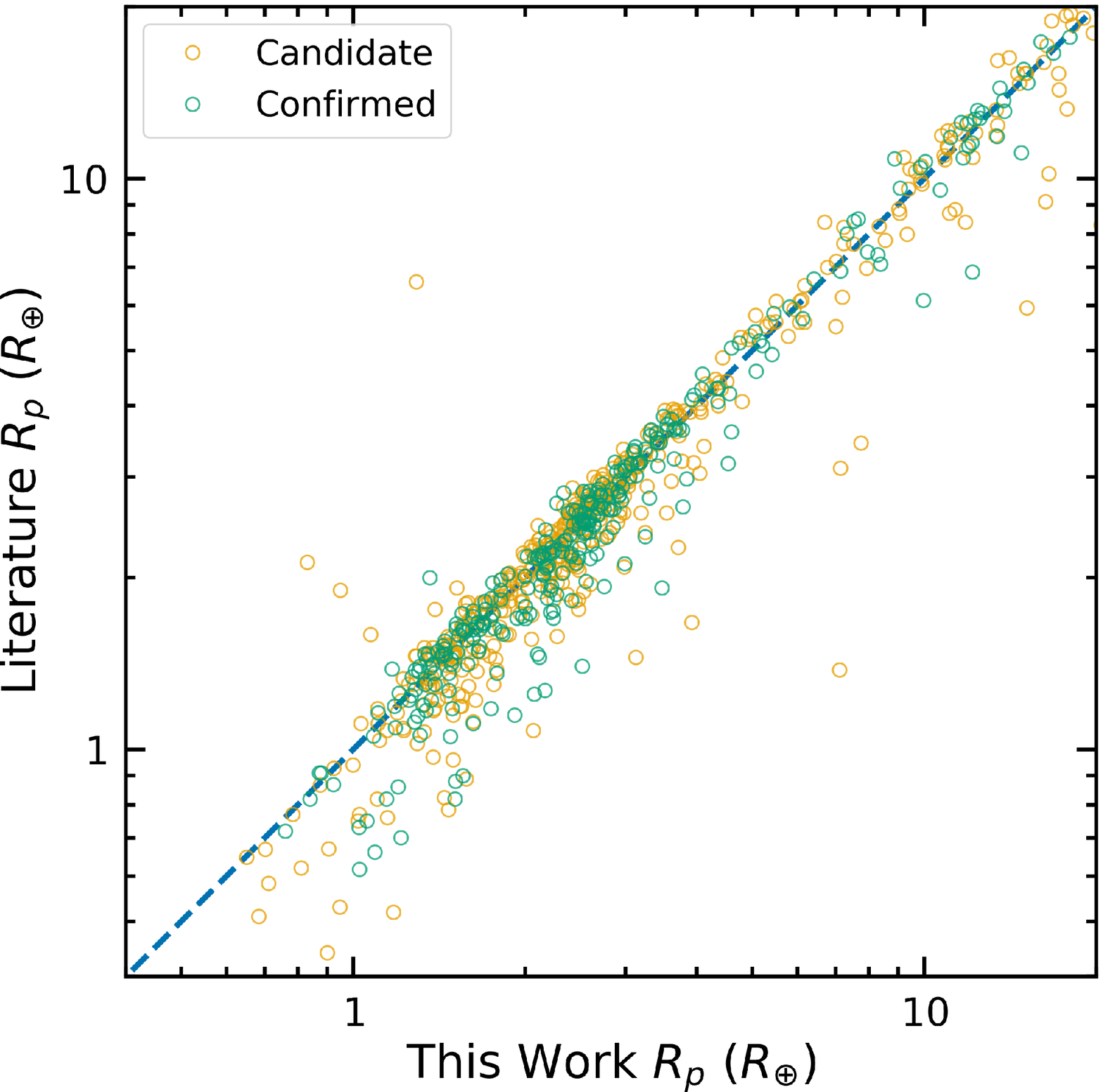}{0.436\textwidth}{\ \ \ \ \ \ (b)}}
    \caption{Comparison of literature stellar radii (a) and planet radii (b) versus our measurements for confirmed and candidate hosts. Our radii for stars smaller than the Sun are typically larger than the values in the literature.}\label{fig:17}
\end{figure*}

\subsection{\kt\ planet hosts and the planet radius valley}\label{sec:gap}

We also compared $R_{\star}$ measurements for candidate and confirmed planet hosts\footnote{\href{https://exoplanetarchive.ipac.caltech.edu/cgi-bin/TblView/nph-tblView?app=ExoTbls\&config=k2candidates}{https://exoplanetarchive.ipac.caltech.edu/cgi-bin/TblView/nph-tblView?app=ExoTbls\&config=k2candidates}}, using the most recent measurements from the literature for targets with previously measured $R_{\star}$ and $R_p/R_{\star}$ (Figure~\ref{fig:17}a). This yielded parameters for 517 candidate and 299 confirmed planets and their hosts for which we also had an $R_{\star}$ measurement. We do not have new parameters for 375 candidates and 93 confirmed planets, which is due to either lack of previously measured $R_{\star}$ and $R_p/R_{\star}$ from the literature, lack of \gaia\ parallaxes, or the planet hosts do not fall within the color space necessary for our classification. For stars with radii less than $5\,R_{\odot}$, our $R_{\star}$ measurements are on average 8.6\% and 7.9\% larger than literature values for candidate and confirmed planet hosts, respectively. Looking specifically at M dwarfs with radii less than $0.6\,R_{\odot}$, our measurements are on average 18.5\% and 33.3\% larger for candidate and confirmed planet hosts. We attribute this significant discrepancy to previous measurements of M dwarf properties using older models which tend to underestimate the radii of cool stars. Using similar measurement techniques, \citet{Hardegree-Ullman2019} and \citet{Dressing2019} also noted that catalog radii for \kepler\ and \kt\ M dwarfs were underestimated by $\sim$40--50\%.

\begin{figure}[ht!]
    \centering
    \includegraphics[width=0.46\textwidth]{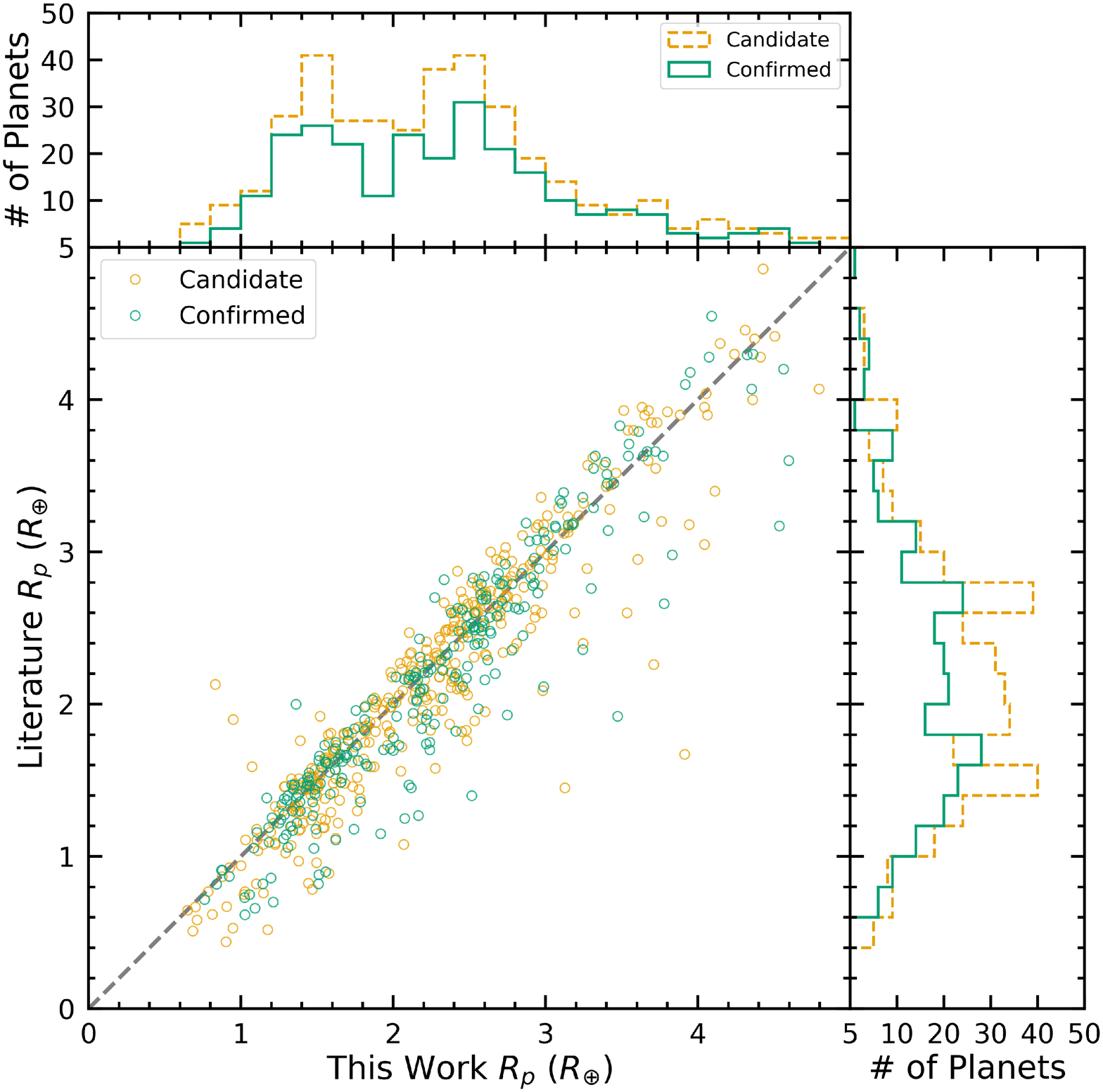}
    \caption{A closer inspection of planets with $R_p<5\,R_{\oplus}$ from Figure~\ref{fig:17}b elucidates a planet radius valley around $R_p\approx1.9\,R_{\oplus}$ using our updated stellar parameters, which was not present in the previously measured planet radii.} \label{fig:18}
		\vspace{-0.5em}
\end{figure}

For proper planet radius measurements, our new stellar properties should be used when fitting the transit light curves to account for effects such as limb darkening on the transit fit. Refitting transit curves is beyond the scope of this paper, but we offer a general quantitative analysis of updated planet radii $R_p$ based on literature values for $R_p/R_{\star}$ and our measurements of $R_{\star}$, which is valid under the assumption that the change of stellar parameters does not significantly affect the measured transit depth. Table~\ref{tab:3} contains our revised planet radii, and Figure~\ref{fig:17}b compares our planet radii to literature values. For planets with $R_p<20\,R_{\oplus}$, our planet radii are on average 6.7\% and 6.8\% larger for candidate and confirmed planets, respectively.

\begin{figure*}[htp!]
    \gridline{\fig{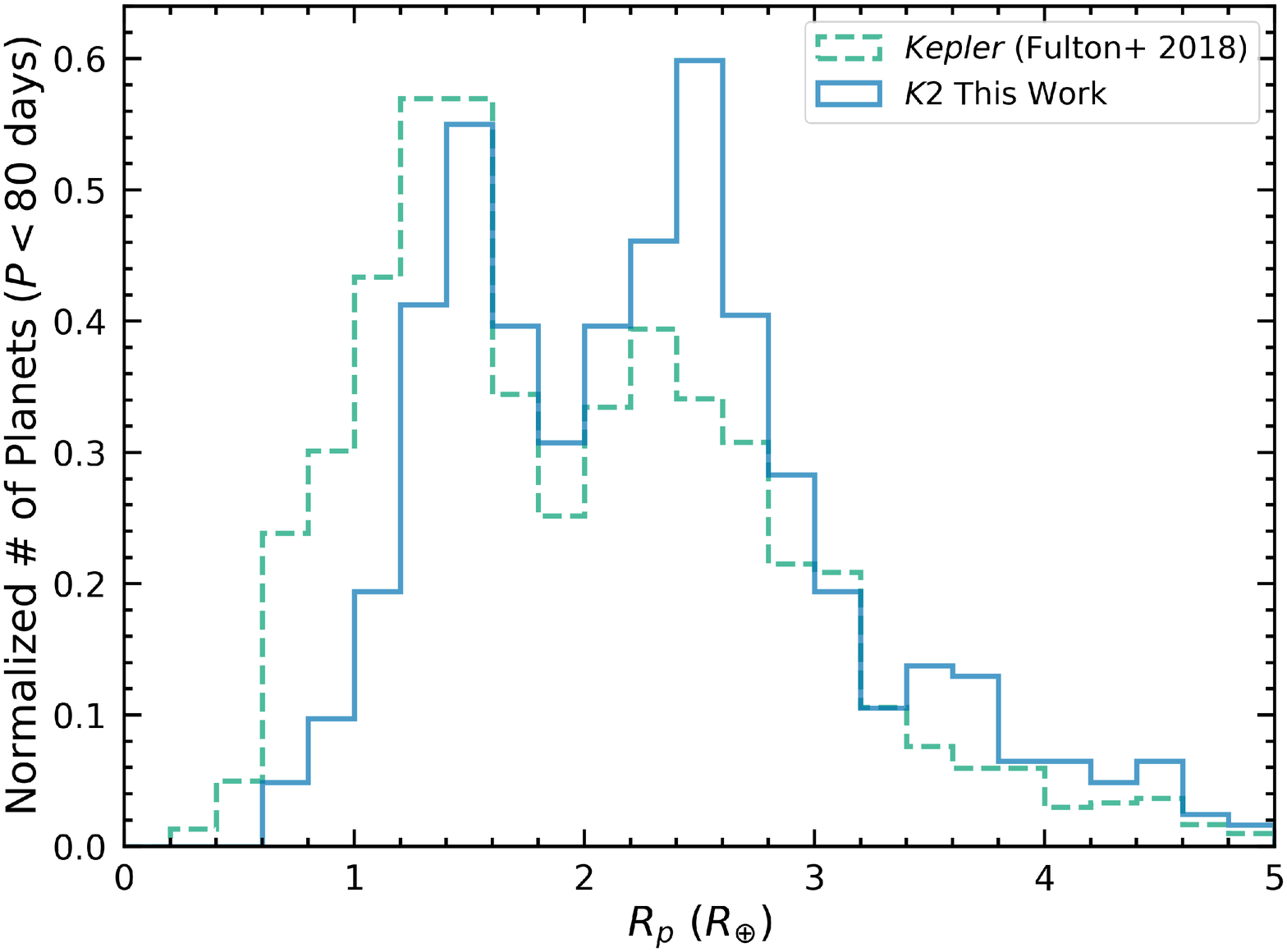}{0.45\textwidth}{\ \ \ \ \ \ (a)}
              \fig{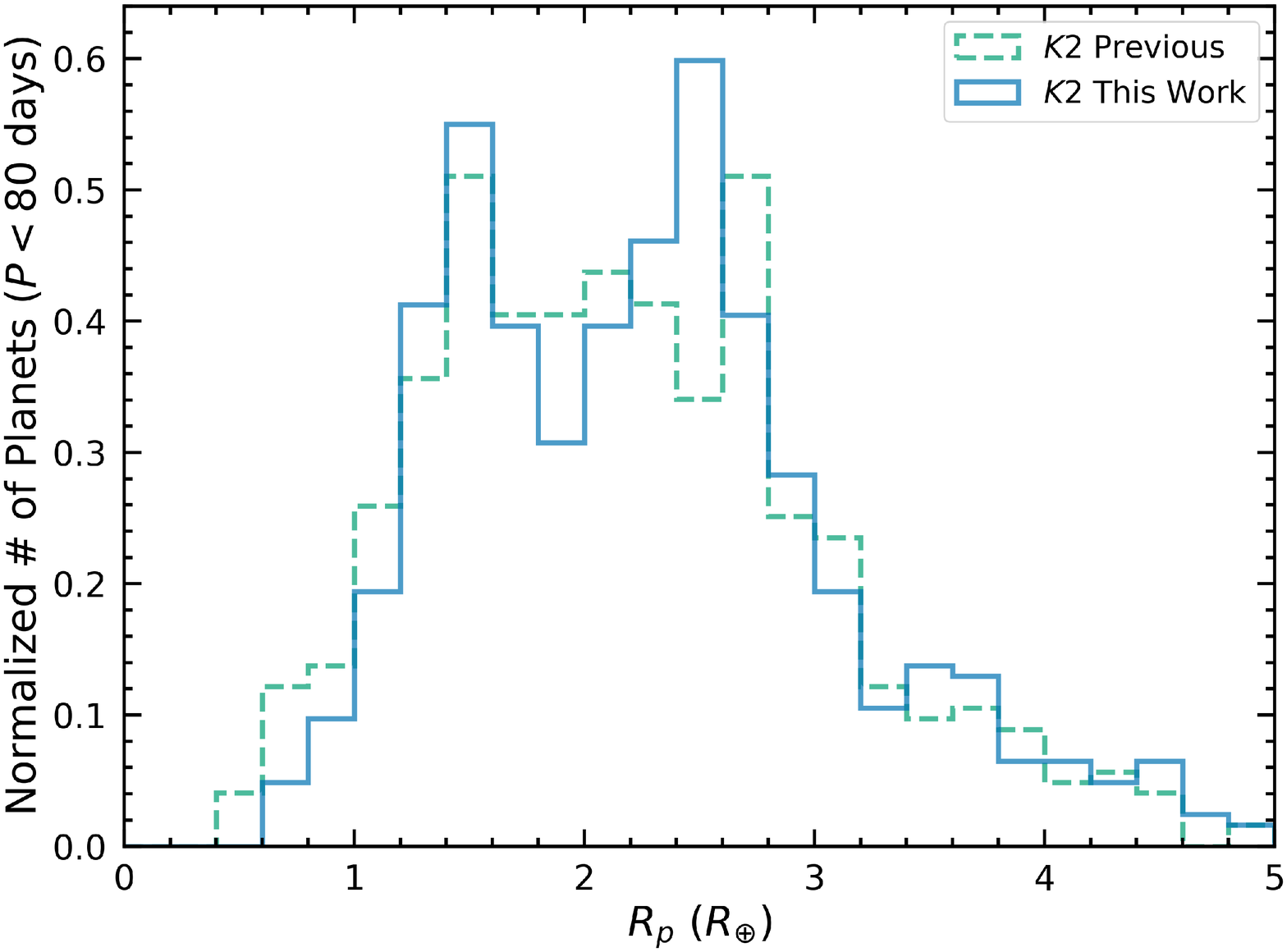}{0.45\textwidth}{\ \ \ \ \ \ (b)}}
    \caption{Normalized planet radius distributions for $R_p<5\,R_{\oplus}$ and $P<80$ days for our combined \kt\ confirmed and candidate sample to the \kepler\ sample from \citet{Fulton2018} (a), and all previous \kt\ measurements (b). There is a much more prominent valley in our measurements than in previous \kt\ measurements. The gap minimum around $R_p\approx1.9\,R_{\oplus}$ that we measure is also consistent with the \kepler\ sample. Note, these measurements have not been corrected for completeness.}\label{fig:19}
\end{figure*}

Taking a closer look at planets with $R_p<5\,R_{\oplus}$, we investigated the planet radius valley, which is not apparent from previous \kt\ planet radii, but is very prominent in our revised radii (Figure~\ref{fig:18}). The \citet{Kruse2019} measurements constitute about 85\% of the previous planet sample, indicating that the differences between our measured stellar radii and the \gaia\ pipeline are not insignificant, and likely due to the differences in $T_{\mathrm{eff}}$. We combined the confirmed and candidate \kt\ planet samples, and compared the planet radius distributions from our measurements to the \kepler\ sample from \citet{Fulton2018} and all previous \kt\ measurements for planets with orbital periods less than 80 days in Figure~\ref{fig:19}. Our updated stellar and planet radii confirm a distinct planet radius valley with a planet sample other than \kepler. This highlights the importance of careful and precise stellar measurements when deriving planet parameters. These measurements were not corrected for completeness, however, which is beyond the scope of this work. Completeness will be addressed in future catalog papers in this series (Zink et al.\ submitted, Zink et al.\ in preparation).

\begin{figure*}[ht!]
    \gridline{\fig{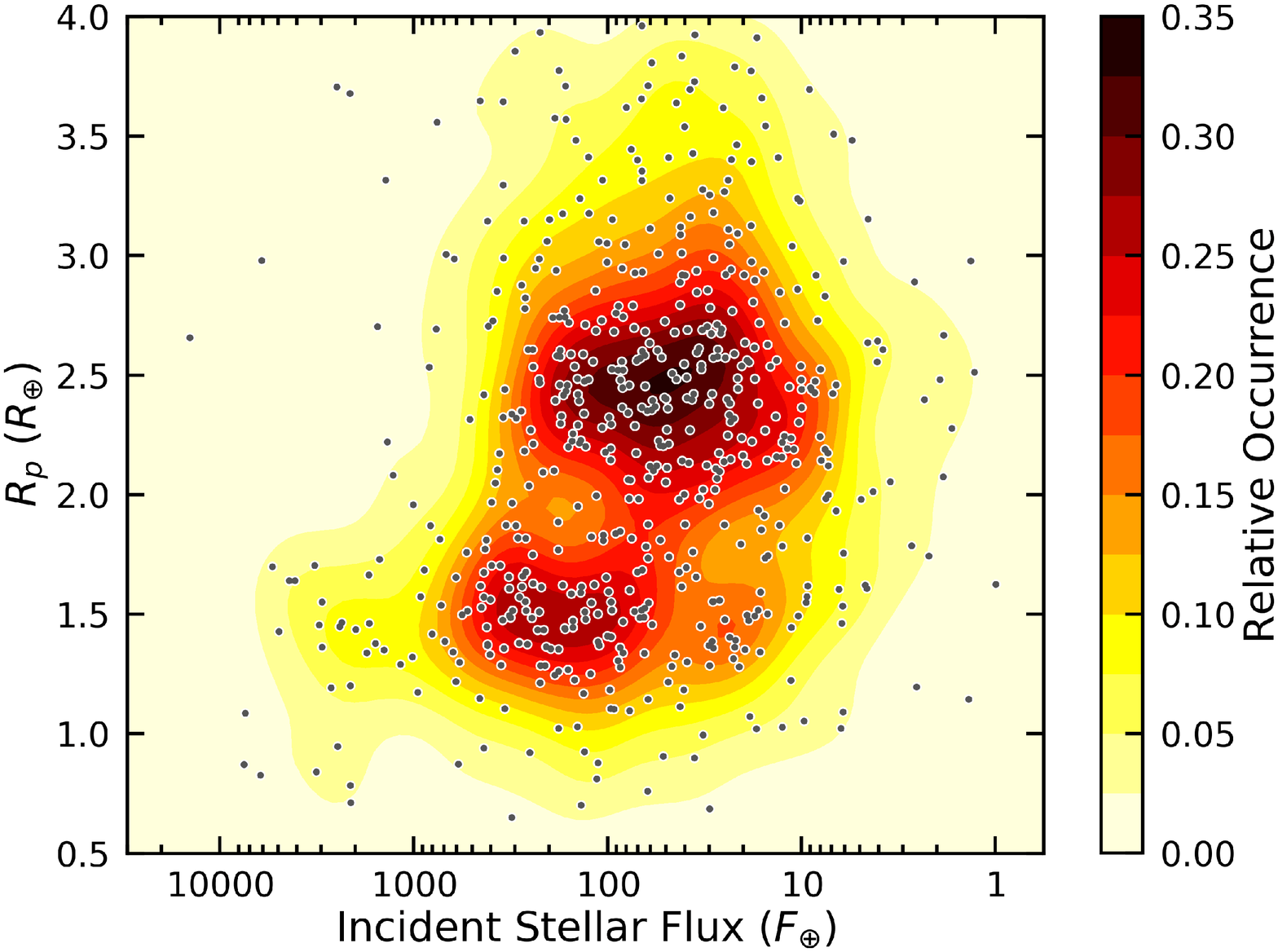}{0.48\textwidth}{(a)}
              \fig{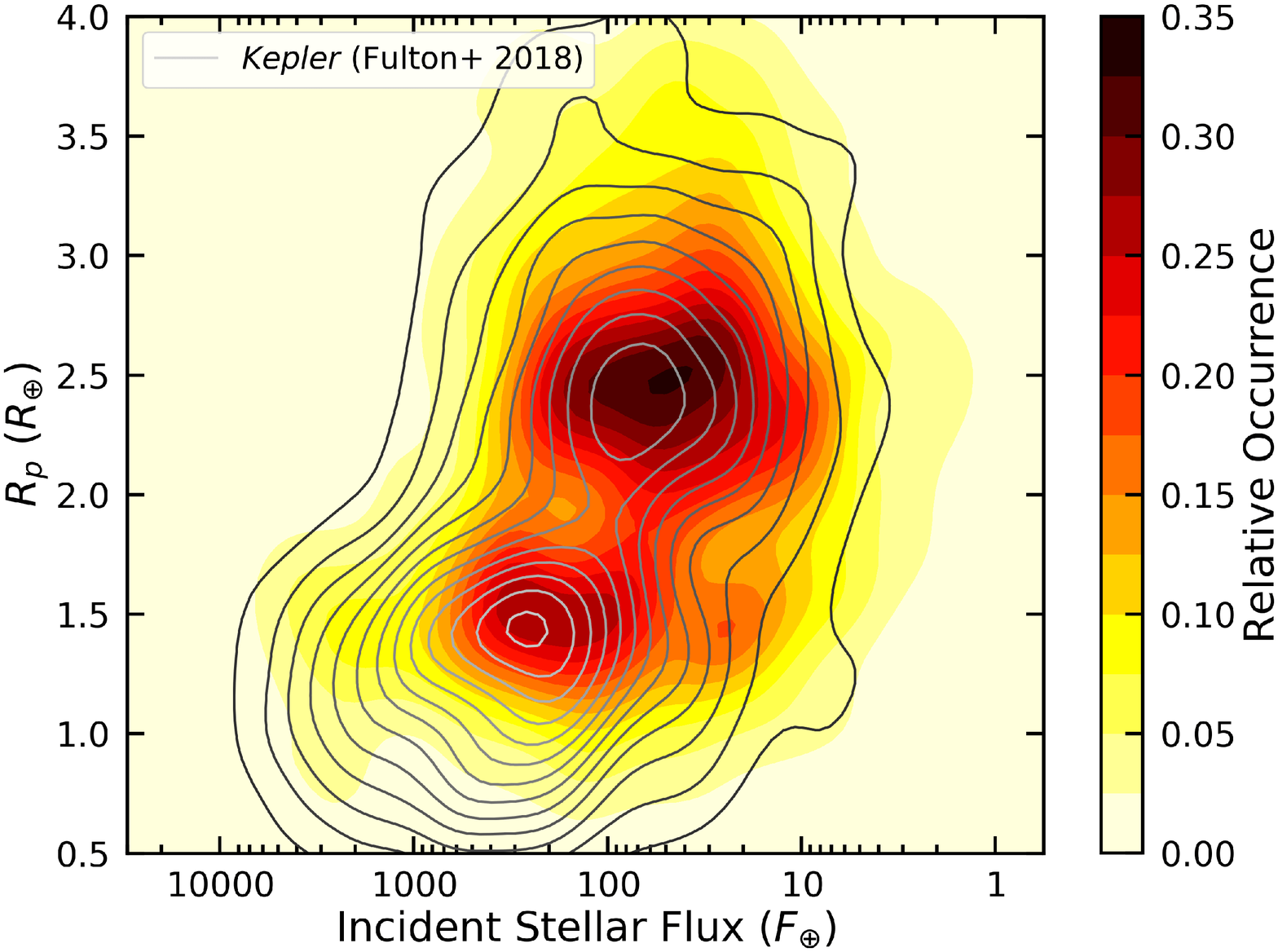}{0.48\textwidth}{(b)}}
    \caption{(a) Planet radius versus incident stellar flux for \kt\ planets. Each point is either a confirmed or candidate planet, and the contours show the density of planets. A valley is present around $2\,R_{\oplus}$, separating a population of super-Earths and sub-Neptunes. (b) The same density contours for \kt\ from (a) with the density contours for \kepler\ planets from \citet{Fulton2018}. The planet populations have similar distributions, with a similar radius gap showing a small slope with respect to incident stellar flux. Note that these plots have not been corrected for completeness.}\label{fig:20}
\end{figure*}

Using the literature values for orbital period and our computed stellar masses, we calculated semi-major axes $a$ for our set of \kt\ planets from Kepler's third law. We then computed incident stellar flux $F_{pl}/F_{\oplus}=(L_{\star}/L_{\odot})(\mathrm{AU}/a)^2$, where stellar luminosity $L_{\star}/L_{\odot}=(R_{\star}/R_{\odot})^{2}(T_{\mathrm{eff}}/T_{\odot})^{4}$ computed using our values. In Figure~\ref{fig:20} we show planet radius versus incident stellar flux for planets smaller than $4\,R_{\oplus}$ and orbital periods shorter than 80 days. The density contours show two relatively distinct populations of planets separated by a valley around $2\,R_{\oplus}$ and a wide range of incident fluxes. As a qualitative comparison, we also show the density contours of the \kt\ planet population and the \kepler\ population from \citet{Fulton2018}. In both cases, the radius valley is apparent at about the same location, with hits of a small slope as a function of incident stellar flux.

\begin{figure*}[ht!]
    \centering
    \includegraphics[width=0.98\textwidth]{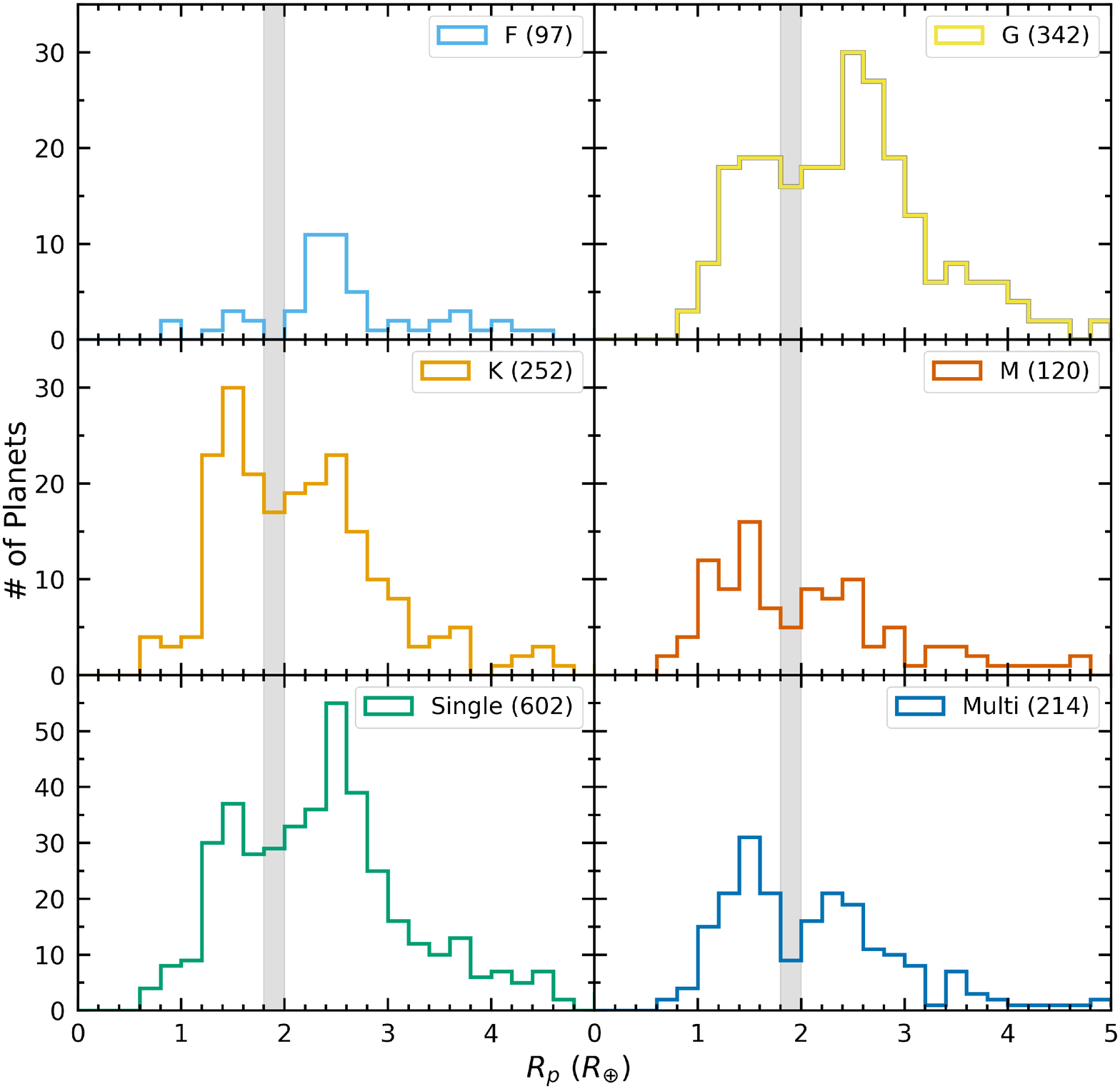}
    \caption{Planet radius distributions separated by spectral type, including both confirmed and candidate planets (top four panels), and single versus multiple planet systems (bottom two panels). We have shaded the region at $1.9\,R_{\oplus}$ for reference in comparison to the radius valley of the total sample. For each spectral type there is evidence for the planet radius valley, which is most prominent for K-type stars. Single planet systems appear to have about twice the fraction of sub-Neptunes compared to super-Earths, whereas the ratio is near unity for multiple planet systems. We again note that these distributions have not been corrected for completeness, so conclusions about planet occurrence rates cannot be drawn from these data.} \label{fig:21} 
		\vspace{-0.25em}
\end{figure*}

\begin{figure*}[ht!]
    \gridline{\fig{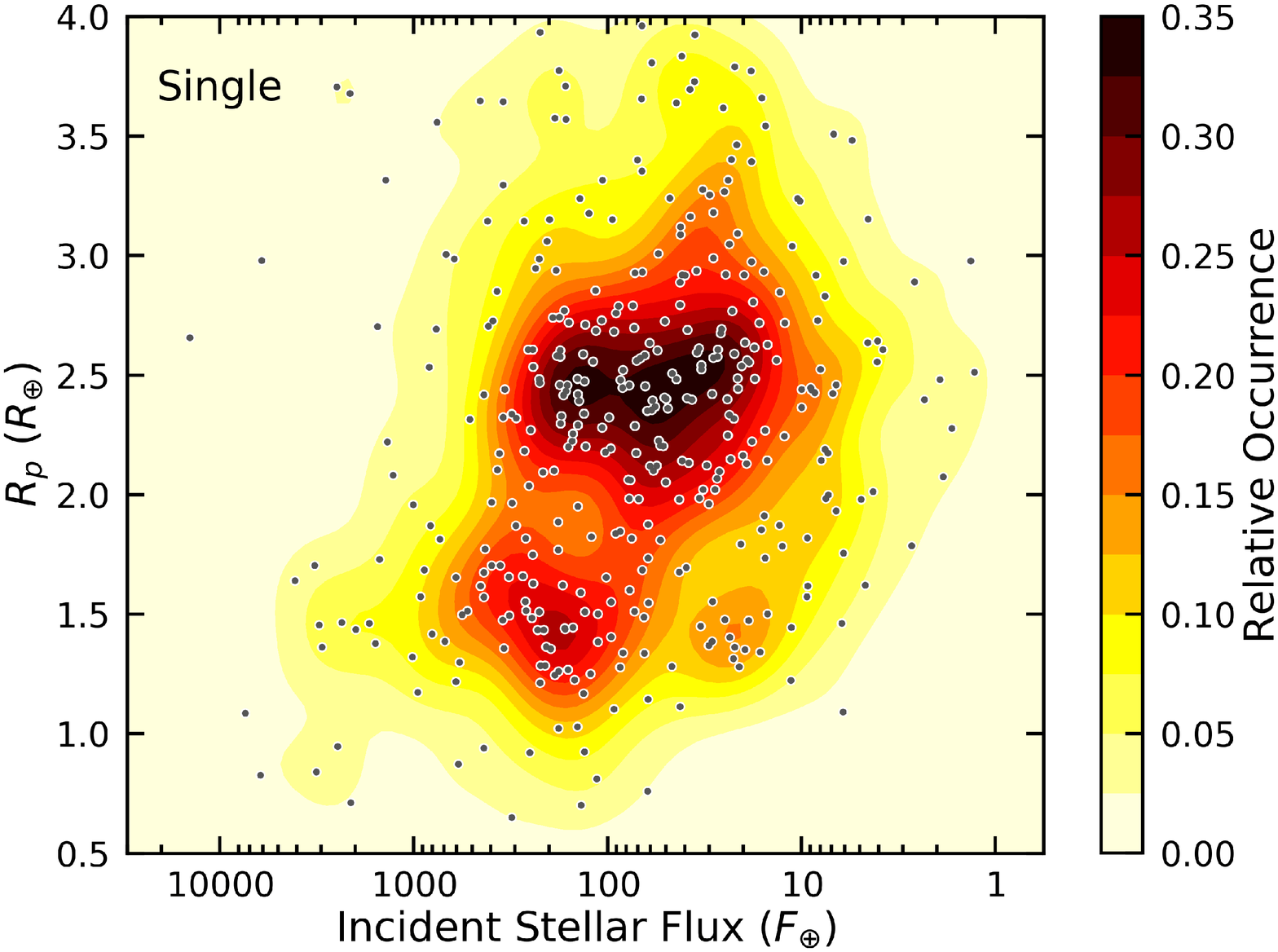}{0.48\textwidth}{(a)}
              \fig{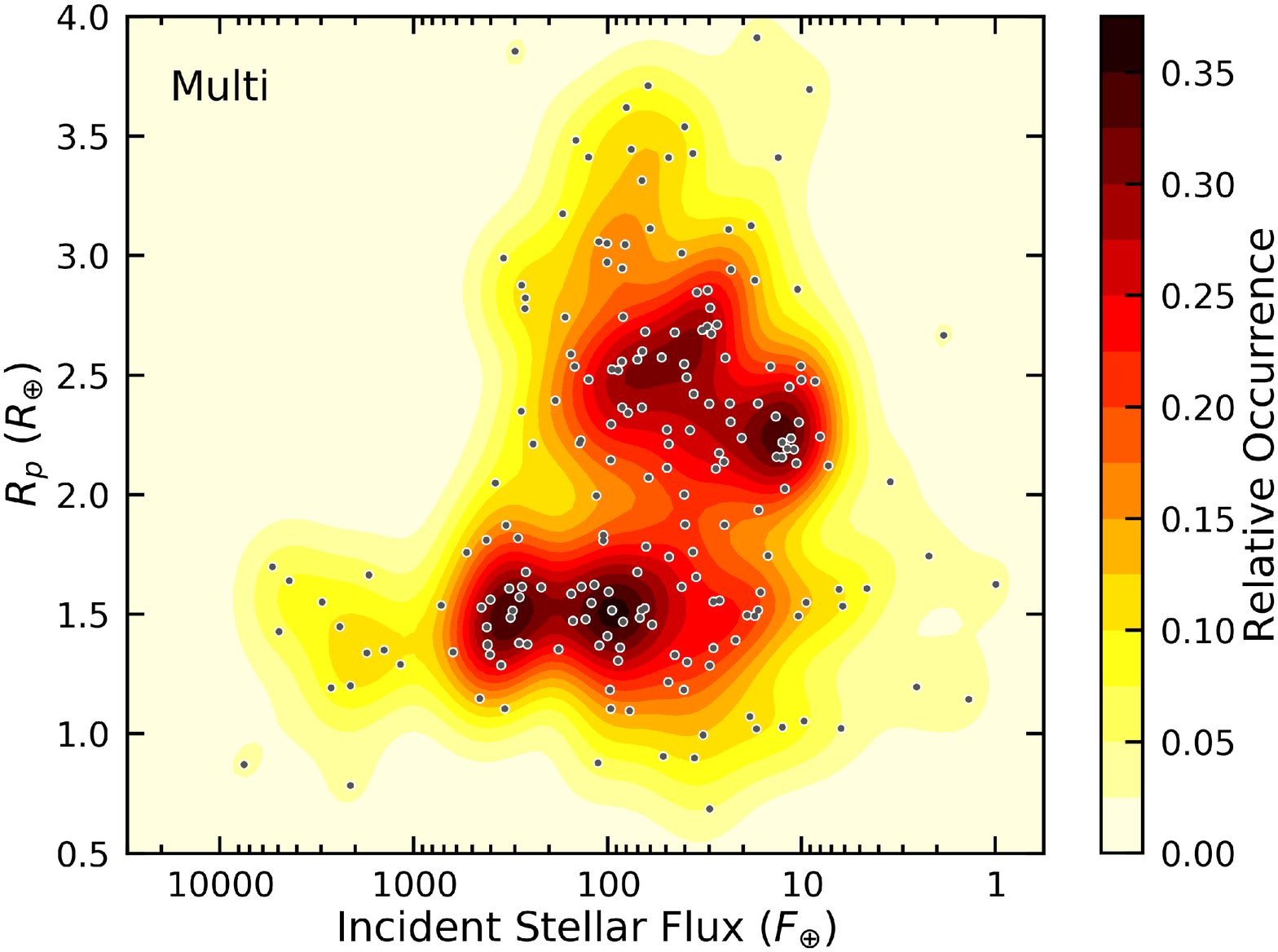}{0.48\textwidth}{(b)}}
    \caption{Planet radius versus incident stellar flux for \kt\ planets in single (a) and multiple planet systems (b). Perhaps more evident than in Figure~\ref{fig:21}, single planet systems appear to have twice as many sub-Neptunes than super-Earths, whereas multiple planet systems have roughly equal numbers of each. Note that these plots have not been corrected for completeness.}\label{fig:22}
\end{figure*}

Since we have spectral types for all of our stars, we separated the \kt\ planet radius distributions by spectral type (Figure~\ref{fig:21}). For each spectral type, there is a lack of planets at $R_p\approx1.9\,R_{\oplus}$. K-type stars show a prominent radius valley, but all other spectral types at least hint at a valley. A larger sample size would be necessary to confirm a valley for F and M stars. Indeed, by combining 275 confirmed \kepler\ and 53 confirmed \kt\ K and M dwarf planets with host star $T_{\mathrm{eff}} < 4,700\,\mathrm{K}$, \citet{Cloutier2019} showed a more definitive planet valley around $1.54\,R_{\oplus}$ for planets around cool stars. Further, there is an increasing total fraction of super-Earths ($R_{\oplus} < R_p < 1.9\,R_{\oplus}$) to sub-Neptunes ($1.9\,R_{\oplus} < R_p < 3.86\,R_{\oplus}$) toward later-type stars, with ratios of 0.20, 0.50, 0.82, and 1.13 for F, G, K, and M stars, respectively, which is consistent with conclusions of planet occurrence rate studies \citep[e.g.,][]{Howard2012,Dressing2015,Mulders2015b,Hardegree-Ullman2019}, indicating that smaller planets are more common toward later spectral types. This effect, however, could be an observational bias, since it is more difficult to detect smaller planets around larger stars.  We also compared the planet radius distributions for single and multiple planet systems in Figures~\ref{fig:21} and \ref{fig:22}. There are 602 single planet systems and 90 multiple planet systems containing a total of 214 planets. For single planet systems, the ratio of super-Earths to sub-Neptunes is 0.51, whereas for multiple planet systems the ratio is 1.02. We leave the analysis of these effects to future studies.

\subsection{Future directions}\label{sec:future}
Our uniformly derived catalog of updated stellar parameters for 222,088 \kt\ stars using LAMOST spectra, \gaia\ parallaxes, and photometry is a crucial step in the process of calculating \kt\ planet occurrence rates. All of the planet candidates analyzed in this paper were from \kt\ Campaigns 1--13, since catalogs for those planets have already been made and are available on the Exoplanet Archive. The next step toward computing planet occurrence rates is to develop a pipeline to uniformly process \kt\ light curves and automatically identify and vet planet candidates across all campaigns (Zink et al.\ submitted, Zink et al.\ in preparation). This will enable us to conduct crucial completeness and reliability tests necessary for accurate planet occurrence rate calculations, and which we have not been able to account for in this work. With a larger set of planet candidates across all campaigns, a more complete analysis of effects such as the planet radius gap can be assessed. Our large set of $T_{\mathrm{eff}}$, $\log\,g$, [Fe/H], $R_{\star}$, and $M_{\star}$ can also enable other statistical population studies of stars and planets.

In this study we have largely ignored the effects of stellar multiplicity. \citet{Duchene2013} estimate that 44\% of all FGK stars are part of a multiple stellar system, and \citet{Winters2019} found a multiplicity rate of $\sim$27\% for M dwarfs within 25\,pc of the Sun. \gaia\ is able to resolve binary stars of similar brightness with separations down to about one arcsecond\footnote{\href{https://www.cosmos.esa.int/web/gaia/science-performance}{https://www.cosmos.esa.int/web/gaia/science-performance}}, however, \citet{Horch2014} estimate that 40-50\% of planet candidate systems host a bound binary within one arcsecond. Our stellar parameters assume a single star or a wide separation such that we can resolve our target. If the stars are actually in multiple systems our stellar radii will typically be overestimated, which could have a significant impact on derived planet parameters and conclusions regarding planet populations \citep[e.g.,][]{Ciardi2015,Furlan2017,Horch2017,Matson2018}. High-resolution imaging surveys to determine stellar multiplicity rates have largely focused on stars with planet candidates, but it is possible that there are differences in multiplicity rates for hosts versus non-hosts, which could suggest differences in formation mechanisms. We strongly encourage additional {high-resolution imaging and high-resolution spectroscopic} observations of \kt\ stars, including stars without known planets, that enable us to more effectively mitigate and assess the impact of stellar companions on planet occurrence rates.

\section{Acknowledgements}

We thank the anonymous referee who provided several helpful suggestions that improved this manuscript. We would also like to thank Christina Hedges, Geert Barentsen, and Jessie Dotson at the \kepler\ Guest Observer office for constructive discussions about \kt\ and this work.

This research has made use of the NASA Exoplanet Archive, which is operated by the California Institute of Technology, under contract with the National Aeronautics and Space Administration under the Exoplanet Exploration Program.

The Pan-STARRS1 Surveys (PS1) and the PS1 public science archive have been made possible through contributions by the Institute for Astronomy, the University of Hawaii, the Pan-STARRS Project Office, the Max-Planck Society and its participating institutes, the Max Planck Institute for Astronomy, Heidelberg and the Max Planck Institute for Extraterrestrial Physics, Garching, The Johns Hopkins University, Durham University, the University of Edinburgh, the Queen's University Belfast, the Harvard-Smithsonian Center for Astrophysics, the Las Cumbres Observatory Global Telescope Network Incorporated, the National Central University of Taiwan, the Space Telescope Science Institute, the National Aeronautics and Space Administration under Grant No. NNX08AR22G issued through the Planetary Science Division of the NASA Science Mission Directorate, the National Science Foundation Grant No. AST-1238877, the University of Maryland, Eotvos Lorand University (ELTE), the Los Alamos National Laboratory, and the Gordon and Betty Moore Foundation.

This work has made use of data from the European Space Agency (ESA) mission {\it Gaia} (\url{https://www.cosmos.esa.int/gaia}), processed by the {\it Gaia} Data Processing and Analysis Consortium (DPAC, \url{https://www.cosmos.esa.int/web/gaia/dpac/consortium}). Funding for the DPAC has been provided by national institutions, in particular the institutions participating in the {\it Gaia} Multilateral Agreement.

This work made use of the \href{http://gaia-kepler.fun}{gaia-kepler.fun} cross-match database created by Megan Bedell. This research made use of the cross-match service provided by CDS, Strasbourg.

Guoshoujing Telescope (the Large Sky Area Multi-Object Fiber Spectroscopic Telescope, LAMOST) is a National Major Scientific Project built by the Chinese Academy of Sciences. Funding for the project has been provided by the National Development and Reform Commission. LAMOST is operated and managed by the National Astronomical Observatories, Chinese Academy of Sciences.

K.\,H-U acknowledges funding from NASA ADAP grant 80NSSC18K0431.

\facilities{Exoplanet Archive, \gaia, \kepler, LAMOST, PS1}

\software{\textsf{astropy} \citep{astropy2013,astropy2018}, \textsf{dustmaps} \citep{Green2018}, \textsf{iPython} \citep{Perez2007}, \textsf{M\_-M\_K-} \citep{Mann2019}, \textsf{matplotlib} \citep{Hunter2007}, \textsf{numpy} \citep{Oliphant2015}, \textsf{pandas} \citep{McKinney2010}, \textsf{scikit-learn} \citep{Pedregosa2011}, \textsf{scipy} \citep{Jones2001}, \textsf{SpectRes} \citep{Carnall2017}}

\newpage
\bibliography{bib}\setlength{\itemsep}{-2mm}
\bibliographystyle{aasjournal}

\newpage
\begin{deluxetable*}{ccccccc}

\tablecaption{\emph{K2} stellar parameters.}\label{tab:1}

\tablehead{
\colhead{EPIC ID} & \colhead{K2 Campaign} & \colhead{Pan-STARRS ID} & \colhead{Gaia DR2 ID} & \colhead{LAMOST ID} & \colhead{$m_g$} & \colhead{\nodata} \\
\colhead{} & \colhead{} & \colhead{} & \colhead{} & \colhead{} & \colhead{(mag)} & \colhead{}}

\startdata
201048855 & 10 & \nodata & 3582456140266586240 & \nodata & $12.320\pm0.040$ \\
201049999 & 10 & \nodata & 3582457617736883840 & \nodata & $13.353\pm0.030$ \\
201050049 & 10 & \nodata & 3582457858255051392 & \nodata & $14.043\pm0.040$ \\
201050511 & 10 & \nodata & 3582458579809568256 & \nodata & $11.405\pm0.030$ \\
201051317 & 10 & 98321820422274493 & 3582459163925111552 & \nodata & $15.285\pm0.003$ \\
201051625 & 10 & \nodata & 3582459301364064768 & \nodata & $12.821\pm0.040$ \\
201052484 & 10 & \nodata & 3582465176879327488 & \nodata & $13.713\pm0.050$ \\
201054099 & 10 & 98441820384293565 & 3582468612853166080 & \nodata & $15.090\pm0.003$ \\
201054338 & 10 & \nodata & 3582466619988381184 & \nodata & $12.188\pm0.050$ \\
201054542 & 10 & 98461822154792184 & 3582466997945503616 & \nodata & $14.871\pm0.005$ \\
201054991 & 10 & 98471821477879568 & 3582467582061055360 & \nodata & $15.185\pm0.003$ \\
201071559 & 10 & 99101827692668111 & 3582605914368082816 & \nodata & $16.649\pm0.005$ \\
201071583 & 10 & 99101826522918865 & 3582603406107179264 & \nodata & $18.861\pm0.014$ \\
201071950 & 10 & 99121828753012622 & 3582605502051225216 & \nodata & $14.848\pm0.003$ \\
201071997 & 10 & 99121826550054361 & 3582603440466918016 & \nodata & $17.159\pm0.004$ \\
201072036 & 10 & \nodata & 3594613577775506048 & \nodata & $15.455\pm0.020$ \\
201072674 & 10 & 99141828915398845 & 3582607220038146176 & \nodata & $14.921\pm0.002$ \\
201073202 & 10 & \nodata & 3594613440336532224 & \nodata & $15.761\pm0.030$ \\
201073315 & 10 & 99171829021981565 & 3582607323117362688 & \nodata & $15.693\pm0.060$ \\
201073427 & 10 & \nodata & 3594616154755833984 & \nodata & $14.027\pm0.020$ \\
201073453 & 10 & \nodata & 3594616253538685824 & \nodata & $13.622\pm0.040$ \\
201073867 & 10 & 99191826959622285 & 3582610346774334336 & \nodata & $17.607\pm0.006$ \\
201073911 & 10 & 99191829080563936 & 3582607421901033856 & \nodata & $16.277\pm0.006$ \\
201074123 & 10 & 99201827274061451 & 3582610003176952320 & \nodata & $17.683\pm0.002$ \\
201074212 & 10 & \nodata & 3582609865738000000 & \nodata & $11.700\pm0.030$ \\
201074534 & 10 & 99211822135945395 & 3594605709395368832 & \nodata & $15.479\pm0.004$ \\
201074673 & 10 & \nodata & 3594614608567639808 & \nodata & $14.109\pm0.020$ \\
201074674 & 10 & \nodata & 3594605812474584704 & \nodata & $12.950\pm0.030$ \\
201074775 & 10 & \nodata & 3582607834216399104 & \nodata & $12.096\pm0.030$ \\
201074882 & 10 & 99221824760577057 & 3594618010181737088 & \nodata & $14.993\pm0.003$ \\
201075355 & 10 & 99241827756634091 & 3582611652444397312 & \nodata & $18.450\pm0.007$ \\
201075442 & 10 & \nodata & 3594606774547212160 & \nodata & $12.263\pm0.030$ \\
\enddata

\tablecomments{This table is available in its entirety in machine-readable form online.}
\tablecomments{There are 222,088 unique targets in this table. There were 19,829 targets observed in two or three campaigns, which we list as separate entries for each \kt\ campaign. This table contains a total of 244,337 entries.}
\tablecomments{Apparent $g$, $r$, and $i$-band magnitudes are from Pan-STARRS for targets with a Pan-STARRS ID and from UCAC4 or SDSS as reported in the EPIC \citep{Huber2016} otherwise.}
\tablecomments{Spectral type, $T_{\mathrm{eff}}$, $\log\,g$, and [Fe/H] for stars with a LAMOST ID were derived using LAMOST spectra. These parameters for stars without a LAMOST ID were derived using photometry trained on the spectroscopic sample.}

\end{deluxetable*}

\begin{deluxetable*}{cc|cc|cc|cc}

\tabletypesize{\normalsize}

\tablecaption{Number of targets in our LAMOST sample with each spectral type classification.}\label{tab:2}

\tablehead{\colhead{Type} & \colhead{\#} & \colhead{Type} & \colhead{\#} & \colhead{Type} & \colhead{\#} & \colhead{Type} & \colhead{\#}} 

\startdata
A1	&	6	&	F3	&	179	&	G3	&	2399	&	K3	&	682	\\
A2	&	2	&	F4	&	131	&	G4	&	580	&	K4	&	280	\\
A3	&	7	&	F5	&	1649	&	G5	&	4009	&	K5	&	457	\\
A5	&	23	&	F6	&	639	&	G6	&	669	&	K7	&	245	\\
A6	&	33	&	F7	&	1038	&	G7	&	1762	&	M0	&	278	\\
A7	&	109	&	F8	&	276	&	G8	&	1266	&	M1	&	496	\\
A8	&	10	&	F9	&	2122	&	G9	&	563	&	M2	&	377	\\
A9	&	13	&	G0	&	915	&	K0	&	363	&	M3	&	195	\\
F0	&	962	&	G1	&	328	&	K1	&	1155	&	M4	&	40	\\
F2	&	703	&	G2	&	1861	&	K2	&	17	&	M5	&	2	\\
\enddata

\end{deluxetable*}

\begin{deluxetable*}{cccccccccccccc}
\tabletypesize{\footnotesize}

\tablecaption{Refined \kt\ planet parameters.}\label{tab:3}

\tablehead{\colhead{EPIC ID} & \colhead{Candidate ID} & \colhead{Confirmed Planet Name} & \colhead{$R_p/R_{\star}$} & \colhead{Period} & \colhead{Reference} & \colhead{Spectral Type} &	\colhead{\nodata} \\
\colhead{} & \colhead{} & \colhead{} & \colhead{} & \colhead{(days)} & \colhead{} & \colhead{} &	\colhead{}}

\startdata
201110617 & 201110617.01 & K2-156 b & $0.01704^{+0.00139}_{-0.00114}$ & $0.813149^{+0.000050}_{-0.000049}$ & 5 & K5 & \\
201111557 & 201111557.01 & \nodata & $0.01692^{+0.00674}_{-0.00148}$ & $2.302368^{+0.000105}_{-0.000103}$ & 5 & K3 & \\
201127519 & 201127519.01 & \nodata & $0.11511^{+0.00492}_{-0.00336}$ & $6.178369^{+0.000195}_{-0.000172}$ & 5 & K3 & \\
201130233 & 201130233.01 & K2-157 b & $0.01105^{+0.00143}_{-0.00097}$ & $0.365257^{+0.000029}_{-0.000029}$ & 5 & G7 & \\
201132684 & 201132684.01 & K2-158 b & $0.02707^{+0.00275}_{-0.00198}$ & $10.062106^{+0.00227}_{-0.002228}$ & 5 & G7 & \\
201152065 & 201152065.01 & \nodata & $0.0226^{+0.0022}_{-0.0055}$ & $10.6966^{+0.002}_{-0.0021}$ & 3 & K5 & \\
201155177 & 201155177.01 & K2-42 b & $0.0313^{+0.0023}_{-0.0047}$ & $6.68851^{+0.00074}_{-0.00075}$ & 3 & K5 & \\
201160662 & 201160662.01 & \nodata & $0.259^{+0.071}_{-0.099}$ & $1.5374115^{+0.0000062}_{-0.0000061}$ & 3 & F6 & \\
201166680 & 201166680.01 & \nodata & $0.01572^{+0.00173}_{-0.00119}$ & $18.10549^{+0.010083}_{-0.012897}$ & 5 & F2 & \\
201176672 & 201176672.01 & \nodata & $0.18^{+0.011}_{-0.011}$ & $79.9999^{+0.0098}_{-0.0098}$ & 2 & K5 & \\
201197348 & 201197348.01 & \nodata & $0.046^{+0.0038}_{-0.0078}$ & $14.9139^{+0.0018}_{-0.002}$ & 3 & K5 & \\
201205469 & 201205469.01 & K2-43 b & $0.0775^{+0.0034}_{-0.0063}$ & $3.471136^{+0.000079}_{-0.000079}$ & 3 & M1 & \\
201205469 & 201205469.02 & K2-43 c & $0.0391^{+0.0039}_{-0.0113}$ & $2.19945^{+0.00015}_{-0.00014}$ & 3 & M1 & \\
201208431 & 201208431.01 & K2-4 b & $0.0368^{+0.0015}_{-0.0031}$ & $10.0051^{+0.00044}_{-0.00043}$ & 3 & K7 & \\
201211526 & 201211526.01 & K2-244 b & $0.01698^{+0.00312}_{-0.00127}$ & $21.070201^{+0.002413}_{-0.002267}$ & 5 & G3 & \\
201225286 & 201225286.01 & K2-159 b & $0.02439^{+0.00226}_{-0.00134}$ & $12.421078^{+0.001049}_{-0.001001}$ & 5 & G7 & \\
201227197 & 201227197.01 & K2-160 b & $0.03189^{+0.00171}_{-0.00114}$ & $3.705871^{+0.000074}_{-0.000076}$ & 5 & G4 & \\
201231064 & 201231064.01 & K2-161 b & $0.02184^{+0.00518}_{-0.00181}$ & $9.283188^{+0.002052}_{-0.0023}$ & 5 & G5 & \\
201238110 & 201238110.01 & \nodata & $0.0505^{+0.005}_{-0.0129}$ & $7.90417^{+0.00091}_{-0.00148}$ & 3 & M2 & \\
201238110 & 201238110.02 & EPIC 201238110 b & $0.054^{+0.0034}_{-0.0054}$ & $28.1696^{+0.0038}_{-0.0043}$ & 3 & M2 & \\
201239401 & 201239401.01 & \nodata & $0.025^{+0.0019}_{-0.0039}$ & $0.905655^{+0.000049}_{-0.000050}$ & 3 & M2 & \\
201247497 & 201247497.01 & \nodata & $0.087^{+0.011}_{-0.07}$ & $2.75421^{+0.00012}_{-0.00012}$ & 3 & M0 & \\
201259803 & 201259803.01 & \nodata & $0.1173^{+0.0034}_{-0.0035}$ & $1.684208^{+0.000024}_{-0.000024}$ & 3 & M1 & \\
201264302 & 201264302.01 & \nodata & $0.0253^{+0.0018}_{-0.006}$ & $0.2122013^{+0.0000023}_{-0.0000018}$ & 3 & M3 & \\
201295312 & 201295312.01 & K2-44 b & $0.01775^{+0.00066}_{-0.00165}$ & $5.65621^{+0.00026}_{-0.00027}$ & 3 & G0 & \\
201299088 & 201299088.01 & \nodata & $0.04741^{+0.00197}_{-0.00184}$ & $21.204739^{+0.005348}_{-0.005523}$ & 5 & G8 & \\
201324549 & 201324549.01 & \nodata & $0.089^{+0.022}_{-0.039}$ & $2.519386^{+0.000014}_{-0.000014}$ & 3 & F5 & \\
201338508 & 201338508.01 & K2-5 c & $0.0348^{+0.0031}_{-0.0079}$ & $10.93459^{+0.00088}_{-0.00105}$ & 3 & K7 & \\
201338508 & 201338508.02 & K2-5 b & $0.073^{+0.021}_{-0.039}$ & $5.73649^{+0.00033}_{-0.00034}$ & 3 & K7 & \\
201345483 & 201345483.01 & K2-45 b & $0.1431^{+0.005}_{-0.0044}$ & $1.7292577^{+0.0000049}_{-0.0000050}$ & 3 & K5 & \\
201352100 & 201352100.01 & \nodata & $0.03231^{+0.00186}_{-0.00145}$ & $13.383629^{+0.00076}_{-0.000727}$ & 5 & K1 & \\
201357835 & 201357835.01 & \nodata & $0.03044^{+0.001}_{-0.00079}$ & $11.8951^{+0.0014}_{-0.0017}$ & 7 & F8 & \\
201359834 & 201359834.01 & \nodata & $0.266^{+0.081}_{-0.128}$ & $40.1401^{+0.0012}_{-0.0012}$ & 3 & M1 & \\
201366540 & 201366540.01 & \nodata & $0.0346^{+0.0054}_{-0.0295}$ & $7.4433^{+0.0011}_{-0.0012}$ & 3 & K7 & \\
201367065 & 201367065.01 & K2-3 b & $0.0358^{+0.0012}_{-0.0031}$ & $10.05467^{+0.00011}_{-0.00011}$ & 3 & M1 & \\
201367065 & 201367065.02 & K2-3 c & $0.0291^{+0.0027}_{-0.0027}$ & $24.64671^{+0.00054}_{-0.00053}$ & 3 & M1 & \\
201367065 & 201367065.03 & K2-3 d & $0.0273^{+0.0029}_{-0.0048}$ & $44.5574^{+0.0023}_{-0.0023}$ & 3 & M1 & \\
201384232 & 201384232.01 & K2-6 b & $0.0259^{+0.0013}_{-0.003}$ & $30.9403^{+0.0023}_{-0.0027}$ & 3 & G3 & \\
201390048 & 201390048.01 & K2-162 b & $0.01878^{+0.00206}_{-0.00141}$ & $9.457747^{+0.001401}_{-0.001392}$ & 5 & K5 & \\
201393098 & 201393098.01 & K2-7 b & $0.0247^{+0.0015}_{-0.0036}$ & $28.6911^{+0.0037}_{-0.0042}$ & 3 & G6 & \\
201403446 & 201403446.01 & K2-46 b & $0.01764^{+0.00086}_{-0.0021}$ & $19.153^{+0.002}_{-0.0022}$ & 3 & F6 & \\
201427874 & 201427874.01 & K2-163 b & $0.02966^{+0.00301}_{-0.0014}$ & $6.673117^{+0.000316}_{-0.000303}$ & 5 & K4 & \\
201437844 & 201437844.01 & HD 106315 b & $0.01677^{+0.00101}_{-0.00067}$ & $9.554515^{+0.001256}_{-0.001368}$ & 5 & F4 & \\
\enddata

\tablecomments{This table is available in its entirety in machine-readable form online.}
\tablerefs{References for $R_p/R_{\star}$ and Period}: (1) \citet{Adams2016}, (2) \citet{Crossfield2016}, (3) \citet{Kruse2019}, (4) \citet{Mann2017b}, (5) \citet{Mayo2018}, (6) \citet{Osborn2016}, (7) \citet{Zink2019}.

\end{deluxetable*}

\end{document}